\documentclass[preprint2]{aastex}
\usepackage[latin1]{inputenc}
\usepackage{amsmath}
\usepackage{amsfonts}
\usepackage{amssymb}
\usepackage{cite}
\usepackage{graphicx}
\usepackage{epsfig}
\usepackage[caption=false]{subfig}
\usepackage{lscape}
\usepackage{afterpage}
\usepackage{longtable}
\usepackage{boxhandler}
\usepackage{tablefootnote}
\usepackage[amssymb]{SIunits}
\def\fig{Figure}

\def\Fig{Figure}

\def\Sect{Section}
\def\Sects{Sections}

\def\Tab{Table}

\def\eqn{Equation}

\def\deg{$^{o}\,$}
\def\arcm{$^{\prime}\,$}

\def\mjybm   {${\rm m}$Jy\,beam$^{-1}$}
\shorttitle{ NGC~253: VLBI Obs. $\&$ supernova rate modelling}
\shortauthors{Rampadarath et al.}

\begin{document}

\title{Multi-Epoch Very Long Baseline Interferometric Observations of the Nuclear Starburst Region of NGC~253: Improved modelling of the supernova and star-formation rates}


\author{H. Rampadarath\altaffilmark{1 *}, J. S. Morgan\altaffilmark{1}, E. Lenc\altaffilmark{2,3} and S. J. Tingay\altaffilmark{1}}

\altaffiltext{1}{International Centre for Radio Astronomy Research, Curtin University, GPO Box U1987, Perth, WA,
Australia}
\altaffiltext{2}{Sydney Institute for Astronomy, School of Physics, The University of Sydney, NSW, Australia}
\altaffiltext{3}{ARC Centre of Excellence for All-sky Astrophysics (CAASTRO)}
\altaffiltext{*}{hayden.rampadarath@icrar.org}

\date{}

\begin{abstract}

The results of multi-epoch observations of the southern starburst galaxy, NGC~253, with the Australian Long Baseline Array (LBA) at 2.3~GHz are presented.  As with previous radio interferometric observations of this galaxy, no new sources were discovered. By combining the results of this survey with Very Large Array observations at higher frequencies from the literature, spectra were derived and a free-free absorption model was fitted of 20 known sources in NGC~253. The results were found to be consistent with previous studies. The supernova remnant, 5.48$-$43.3, was imaged with the highest sensitivity and resolution to date, revealing a two-lobed morphology. Comparisons with previous observations of similar resolution give an upper limit of $10^{4}$~km~s$^{-1}$ for the expansion speed of this remnant. We derive a supernova rate of $<$0.2~yr$^{-1}$ for the inner 300~pc using a model that improves on previous methods by incorporating an improved radio supernova peak luminosity distribution and by making use of  multi-wavelength radio data spanning 21 years. A star formation rate of $SFR(M\geq5~M_{\odot})<4.9~M_{\odot}$~yr$^{-1}$ was also estimated using the standard relation between supernova and star-formation rates. Our improved estimates of supernova and star-formation rates are consistent with studies at other wavelengths. The results of our study point to the possible existence of a small population of undetected supernova remnants, suggesting a low rate of radio supernovae production in NGC~253.

\end{abstract}


%\maketitle
%-------------------------------------------------------------------------------------------------------------------------------------------------------------------------------------
\section{Introduction}
\label{sec:Intro}

Starburst galaxies are defined as galaxies that are currently undergoing a period of intense star-formation, mainly within (but not exclusive to) the nuclear region, at a rate that cannot be maintained over their lifetime \citep{Weedman1987, LH96}. The radio emission from such galaxies is not dominated by an active supermassive black hole, but by thermal and non-thermal emission that traces the star-forming activity  (e.g. \citealt{Mcdonaldetal02}). The thermal emission is the result of  free-free (or thermal bremsstrahlung) radiation from gas ionised by hot, young stars, while the non-thermal emission is from synchrotron radiation due to the acceleration of electrons to relativistic speeds by the core-collapse of massive stars to form Type~II or Type~Ibc supernovae \citep{Pedlaretal1999, Weileretal2002, Mcdonaldetal02}. Interaction of the supernova with the local circumstellar medium generates a hot, shocked region that produces a prompt emission typically observed as a radio supernova (RSN) \citep{Weileretal2002, CFN06}. As the RSN expand into the surrounding interstellar medium, they evolve into supernova remnants (SNRs), forming shells that emit non-thermal synchrotron radiation that can be visible over many years \citep{Woltjer1972,UA97, FerreiraandJager2008}.  

Studies of SNRs within nearby starbust galaxies can provide important information on the astrophysical processes occurring within these galaxies. Obscuration by dust and gas limit observations at shorter wavelengths, while radio wavelengths are largely unaffected. In addition, high-resolution, wide-field radio interferometry offers an opportunity to detect and resolve individual SNRs (e.g. \citealt{Mcdonaldetal02, Tingay04,  LT06, LT09, Fenechetal10}). Such observations over multiple epochs allow monitoring of the evolution of existing SNRs and the detection of new RSNe and/or SNRs, enabling the supernova and star formation history of the galaxy to be reconstructed. Such studies have been successfully applied to nearby starburst galaxies such as: M82 \citep{Pedlaretal1999, Mcdonaldetal02, Beswicketal06, Fenechetal10, Gendreetal2012}; Arp~220 \citep{Rovilosetal2005, Lonsdaleetal2006, Paraetal2007, Batejatetal2012}; Arp~299 \citep{Ulvestad2009, Canizalesetal11, Bondietal2012}; NGC~4945 \citep{LT09}; and NGC~253 \citep{UA97, Tingay04, LT06}. These studies have provided a wealth of information concerning RSNe and the evolution of SNRs, the star-formation and supernova rates of the host galaxies, and the interstellar medium of the starburst region. For example, almost three decades of observations have identified $\sim$100 compact sources in M82 \citep{Fenechetal10}, where most have been resolved into parsec-scale shell-like structures \citep{Muxlowetal1994, Beswicketal06, Fenechetal10}. In addition, these multi-epoch observations have been instrumental in measuring the expansion speeds of the resolved SNRs (2,000 - 11,000~kms$^{-1}$), and with giving a direct estimate of the supernova rate (0.09~yr$^{-1}$) in M82. The discovery of new sources such as SN2008iz \citep{Brunthaleretal2009b} and the radio transient 43.78+59.3 \citep{Muxlowetal2010} would not have been possible without regular radio monitoring of M82. In addition, by combining multi-wavelength, high resolution radio observations of 46 compact sources in M82, \citet{Mcdonaldetal02} were able to identify low-frequency spectral turnovers, due to free-free absorption by the surrounding ionized medium.

As one of the nearest star-forming galaxies, NGC~253 has been extensively studied from gamma-rays to radio wavelengths. Recently, as part of the Advanced Camera for Surveys (ACS) Nearby Galaxy Survey Treasury (ANGST), \citet{Dalcantonetal09} estimated the distance to NGC~253 as 3.47~$\pm$~0.24~Mpc, 3.46~$\pm$~0.07~Mpc and 3.40~$\pm$~0.09~Mpc. The weighted average of the \citet{Dalcantonetal09} distance estimates, 3.44~$\pm$~0.13~Mpc, is used as the distance to NGC~253 in this paper.

The first high-resolution, wide-field radio interferometric observations of NGC~253 were made by \citet[hereafter TH85]{TH85} with the Very Large Array (VLA) at 15~GHz (2~cm). With a resolution of $0^{\prime\prime}.21~\times~0^{\prime\prime}.10$, TH85 discovered nine compact sources (designated TH1 to TH9) within the central nuclear region. Following the results of TH85, Ulvestad and Antonucci conducted almost a decade of multi-frequency radio (1.8, 5, 8.3, 15 and 23~GHz) observations of the nuclear star-forming region of NGC~253 with the VLA, that culminated in a series of papers \citep{UA88, UA91, UA94, UA97}; they identified 64 individual compact sources, including the original nine from TH85. 
Spectral index measurements were obtained for the 17 brightest sources between frequency pairs of 5/15 and 8.3/23 GHz. Almost half of the 17 sources were identified with thermal H\textsc{ii} regions while the remaining sources were taken to be associated with SNRs. Over the course of their survey, no new RSNe or SNRs appeared and the radio flux densities of the detected SNRs were found to be stable \citep[hereafter UA97]{UA97}. This led to an estimate for the upper limit on the supernova rate of 0.3~yr$^{-1}$ by UA97.

In an attempt to resolve the low frequency radio emission in the inner 300~pc, \citet{Tingay04} conducted the first wide-field VLBI observations of NGC~253. The observations were carried out with the Australian Long Baseline Array (LBA) at 1.4~GHz and matched the angular resolution of the UA97 23~GHz VLA observations. While the observations detected only two sources (TH7 and TH9), \citet{Tingay04} showed that the radio emission at low frequencies is absorbed by ionized gas with a free-free optical depth range at 1~GHz of $\tau_{0}~\backsimeq$~2.5 to $>$~8. This result was consistent with observations between 0.33 - 1.5 GHz with the VLA \citep{Carilli1996} and radio-recombination line (RRL) modelling of the nuclear region of NGC~253 \citep{Mohanetal2002}.

Motivated by these indicators of free-free absorption in NGC~253, \citet[hereafter LT06]{LT06} began a program to observe the nuclear region of this  galaxy with the LBA. Their observations, conducted in 2004 at 2.3~GHz, were higher in sensitivity and resolution than \citet{Tingay04} and covered the region observed by UA97. They identified six compact sources, which were also seen in the higher frequency observations with the VLA (UA97). One of the sources, 5.48-43.3 was also resolved into a shell-like structure approximately 90~mas (1.7~pc) in diameter. Combining the LT06 data with the multi-wavelength radio data from UA97, the spectra of 20 compact sources in the nuclear region of NGC~253 were found to be consistent with the free-free absorption interpretation of \citet{Tingay04}. The results indicated that while the free-free optical depth is highest towards the supposed nucleus, it varies significantly ($\tau_{0}~\simeq$~1 to $>$~20) throughout the nuclear region, implying variations in the gas density. Of the 20 sources, eight were found to have flat spectral indices, indicative of thermal  H~\textsc{ii} regions, while the remaining sources were taken to be associated with SNRs due to their steep spectral indices at high frequencies. With no new sources detected in NGC~253 over almost two decades, LT06 developed a Monte-Carlo method based on the work of \citet{UA91}, to estimate the upper limit on the supernova rate. Their model took into consideration improved distance measurements, a median free-free opacity, and the sensitivity limits of six observations over a 17~yr period. A value of 2.4~yr$^{-1}$ was derived for the upper limit on the supernova rate. This high value suggests that there may be a large number of undetected RSNe, with observations over that period only detecting the rare, bright events. Detecting weaker or short-lived SNRs would provide tighter constraints on the supernova rate. However, this can only be done through frequent, high sensitivity, high resolution observations of NGC~253.

In this paper, we present the results of multi-epoch, wide-field VLBI observations conducted at 2.3~GHz with the LBA, of the nuclear region of NGC~253. \Sect~\ref{sec:Obsall} describes the observations, data analysis methods and the sources detected, including cross identifications with previous observations and investigations of possible flux density variations. Free-free absorption modelling of the spectra is described in \Sect~\ref{sec:CS}. The morphology of the resolved SNR, 5.48-43.3, is examined in detail in \Sect~\ref{sec:548}. An improvement to the Monte-Carlo method of LT06 is presented with new estimates of the upper limits on supernova and star-formation rates within the inner 300~pc region of NGC~253 in \Sects~\ref{sec:SR} $\&$ \ref{sec:SFR}. The results are summarised in \Sect~\ref{sec:sum}.

%-----------------------------------------------------------------------------------------------------------------------------------------------------------------------------------
\section{Observations, Data Reduction, and Results}
\label{sec:Obsall}

\subsection{Observations}
\label{sec:Obs}

%--------------------------------------
\begin{center}
\begin{table*}[htp]\scriptsize
\begin{minipage}{18cm}
 \caption{\textsc{2.3~ghz lba multi-epoch observations}}
\label{tab:table1}
\medskip
\hfill{}
\begin{tabular}{lccc}
\hline\hline
\noalign{\smallskip}
 Epoch& 2006 & 2007 & 2008\\
\hline 
Observing Date & May 12/13 & June 22/23 & June 5/6 \\
Array & \textsc{pa~atca~mp~cd~ho} & \textsc{pa~atca~mp~cd~ho} & \textsc{pa~atca~mp~cd~ho~tid}\\
Observing Time (hrs) & 12 & 12 & 10 \\
Frequency Range (MHz) &2269-2333 & 2268-2332 & 2226-2290\\
Bandwidth (MHz) & 64 & 64 & 64 \\
$\#$ of IFs  $\times$ $\#$ of Frequency Channels & 4$\times$16 & 4$\times$16  & 4$\times$16  \\
Polarization	(\textsc{pa~atca~mp})			& RR~LL & RR~LL & RR~LL \\
Polarization	(\textsc{cd~ho~tid})			& RR & RR &  RR \\
Naturally weighted wide-field array rms  (\mjybm) & 0.39 & 0.44 & 0.18 \\
Wide-field array \textsc{clean} beam (mas) & 147$\times$51 @ $-$71\deg & 134$\times$48 @ $-$72\deg & 86$\times$33 @ $-$88\deg\\
Full array \textsc{clean} beam (mas) & 15  $\times$13 @ 0.1\deg & 15$\times$15 @ $-$29\deg & 16$\times$13 @ 37\deg \\
\hline
\end{tabular}
\hfill{}
 \end{minipage}%
\end{table*}
\end{center}

\begin{figure}[htp]
\centering
\includegraphics[angle=-90,scale=0.3]{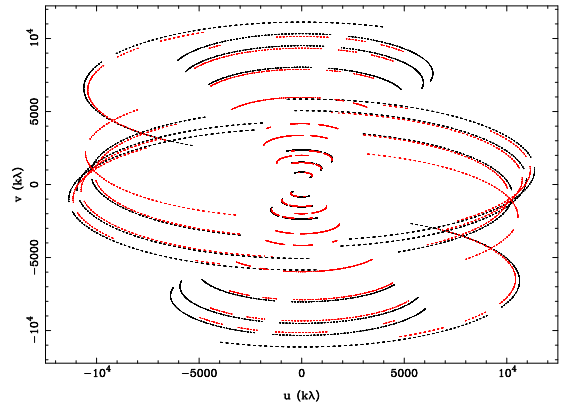}
\caption{$uv$ coverage of the 2007 (black) and 2008 (red) LBA epochs at 2.3~GHz (band centre frequency for each baseline only). The $uv$ coverage for the 2006 epoch (not plotted) was similar to the 2007 $uv$ coverage.}
\label{fig:uvcov}
\end{figure}

%--------------------------------------
%\twocolumn
NGC~253 was observed at 2.3~GHz with the Australian Long Baseline Array (LBA), at three epochs as described in \Tab~\ref{tab:table1}. The observations were carried out with: the 64 m Parkes (Pa) antenna of the Australia Telescope National Facility (ATNF); the ATNF Australia Telescope Compact Array (ATCA)\footnote{5 $\times$ 22 m antennas were used for the 2006 $\&$ 2008 epochs, while 4 $\times$ 22 m antennas were used for the 2007 epoch.}, used as a phased array; the ATNF Mopra (Mp) 22 m antenna; the University of Tasmania's 26 m antenna near Hobart (Ho); and the University of Tasmania's 30 m antenna near Ceduna (Cd). In addition, the 70 m NASA Deep Space Network antenna at Tidbinbilla (Tid) was used for the 2008 epoch. The data for each observation were recorded from 4 $\times$ 16 MHz bands (digitally filtered 2-bit samples) in the frequency ranges given in \Tab~\ref{tab:table1}. All bands had dual circular polarization at Parkes, ATCA, and Mopra, with right circular polarization only at the remaining antennas. \Tab~\ref{tab:table1} lists the parameters associated with the LBA observations. During each observation, 3~minute scans of NGC~253 (centred on: $\alpha$ = 00$^{h}$47$^{m}$ 33.178$^{s}$; $\delta$ =  $-$25$^{\circ}$17$^{\prime}$ 17$^{\prime\prime}$.060 [J2000.0]) were scheduled, alternating with 3 minute scans of a nearby phase reference calibration source, PKS J0038-2459 ($\alpha$ = 00$^{h}$ 38$^{m}$ 14.735$^{s}$; $\delta$ =  $-$24$^{\circ}$ 59$^{\prime}$ 02$^{\prime\prime}$.235 [J2000.0]), located 2.13 degrees from the target. The recorded data for all epochs were correlated using the DiFX software correlator \citep{Delleretal07,Delleretal11}, with an integration time of 2 seconds and 64 frequency channels across each 16 MHz band (channel widths of 0.25 MHz). The $uv$ coverage for the 2007 and 2008 epochs are shown in \fig~\ref{fig:uvcov}.

\bigskip

\subsection{Calibration and Data Reduction}
\label{sec:Calib}

\normalsize

The initial data reduction and calibration were performed using the data reduction package \textsc{aips}\footnote{The Astronomical Image Processing System was developed and is maintained by the National Radio Astronomy Observatory, which is operated by Associated Universities, Inc., under cooperative agreement with the National Science Foundation.}. Prior to calibration, flagging of data during times at which each of the antennas were known to be slewing and time ranges that contained known bad data, were carried out via application of flag files and information provided by the observing logs. Data from the first 30 seconds of each scan from baselines involving ATCA or Parkes were flagged, in order to eliminate known corruption of the data at the start of each scan at these two telescopes.

During correlation, nominal (constant) system temperatures (in Jansky) for each antenna were applied to the correlation coefficients. The nominal calibration was refined by application of antenna system temperatures (in Kelvin) measured during the observation, along with the gain (in Janskys per Kelvin) for each antenna. Further refinements to the amplitude calibration were derived from simultaneously recorded ATCA observations of PKS J0038-2459, which is a bright, compact radio source, unresolved at this frequency on the LBA baselines. Since the LBA isn't resolving out any extended structure, the ATCA and LBA baselines will measure the same flux density. Thus, the flux density measured at the ATCA can be used to check and refine the amplitude calibration for the LBA data. Suitable simultaneous recorded ATCA data were obtained for all but the 2008 observation.

\begin{figure}[htp]
\centering
\includegraphics[scale=0.4]{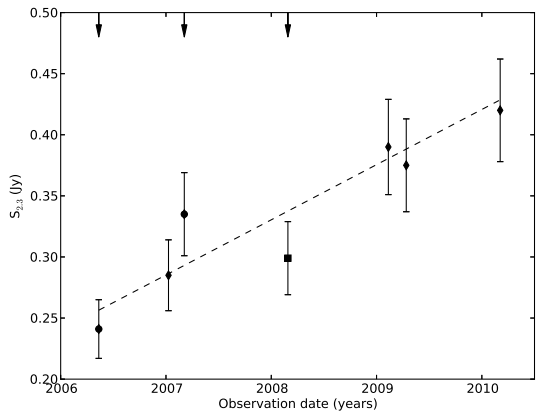} 
\medskip
\caption{Light curve of the PKS~J0038-2459 from 2006 to 2010 at 2.3~GHz. The data are from the Bordeaux VLBI Image Database and the USNO Radio Reference Frame Image Database (points) and the LBA fluxes derived from our 2006 and 2007 observations (circles). A non-linear least-squares fit to the data was used to interpolate the flux density of the calibrator during the 2008 epoch. The resulting fit produced a standard deviation, $\sigma$ = 19 mJy.  The square point plots the final imaged flux density following amplitude refinement for the 2008 epoch (square). The arrows indicate the dates of the LBA observations.}
\label{fig:j0038lightcurve}
\end{figure}

The NRAO's Very Long Baseline Array (VLBA) calibrator survey routinely observes PKS J0038$-$2459 at 2.3~GHz, but did not in 2008. \Fig~\ref{fig:j0038lightcurve} plots the 2.3~GHz light curve of PKS J0038$-$2459 for eight epochs from 1997 to 2010. The error bars are $\pm$10$\%$ of the measured flux densities. A weighted non-linear least-squares fit was used to interpolate to the flux density of PKS J0038$-$2459 at the time of the 2008 LBA observation. The fit gave a flux density of 337~mJy, with a 1$\sigma$ error of 19~mJy at the time of the 2008 observation. This flux density was used to refine the amplitude calibration for the 2008 LBA data, by defining a model with this flux density in \textsc{difmap} \citep{S97}. The \textsc{difmap} task \textsc{gscale} was used to determine corrections to the uncalibrated amplitudes of PKS J0038$-$2459 from the model. 
%The resulting amplitude scaling values for each antenna were then applied to the antenna system temperatures in \textsc{aips} via the \textsc{antab} input file.

%The ATCA flux density measurements used to refine the amplitude calibration for the LBA data (diamonds) and
Following amplitude calibration refinement, global fringe-fitting solutions were determined for PKS J0038$-$2459 (\textsc{aips} task \textsc{fring}) with a three minute solution interval, finding independent solutions for each of the 16 MHz bands. The delay and phase solutions were examined and, following editing of bad solutions, applied to PKS J0038$-$2459. The PKS J0038$-$2459 data were exported to \textsc{difmap}, where the data were vector-averaged over 30~s, flagged of bad data and imaged using standard imaging techniques (deconvolution and self-calibration of both phase and amplitude). The resulting images of PKS J0038-2459 for the 2006, 2007 and 2008 epochs show a highly compact source, with no significant structure on these baselines at this frequency to a dynamic range\footnote{Defined as the ratio of peak of image to peak of brightest artefact.} of 400. The final calibration solutions (phase and amplitude) of PKS J0038$-$2459 were exported via the \textsc{difmap} task, \textsc{cordump}\footnote{http://www.atnf.csiro.au/people/Emil.Lenc/tools/Tools/Cordump.html} \citep{LT09} to an \textsc{aips}-compatible solutions table. The solutions table was then transferred to \textsc{aips} and applied to all sources in the dataset. The PKS J0038-2459 data were also used to derive a bandpass calibration via the \textsc{aips} task \textsc{bpass} which was applied to the NGC~253 data. The edge channels of each band were flagged from the data set (2 channels from both the lower and upper edge of each 16 channel band). The final calibration solutions were applied to both PKS J0038$-$2459 and NGC~253 and the visibility data exported as \textsc{fits} files. 
%------------------------------------------
\afterpage{
%\begin{figure*}[htp]
%\centering
%\begin{tabular}{cc}
%\includegraphics[trim=0cm 2.2cm 0cm 4cm,clip=true,angle=-90,scale=0.48,clip=true]{f3a.eps} &
%\includegraphics[trim=0cm 2.2cm 0cm 0cm,angle=-90,scale=0.48,clip=true]{f3b.eps}\\
%\multicolumn{2}{c}{\includegraphics[trim=0cm 0cm 0cm 0cm,angle=-90,scale=0.48,clip=true]{f3c.eps}}\\\\
%\end{tabular}
%\caption{Wide-field VLBI images of NGC~253 with the LBA at 2.3 GHz. \textit{Top Left: 2006 epoch}. The rms noise level is 0.39~\mjybm. The peak is 10.4~\mjybm  \textit{Top Right: 2007 epoch}. The rms noise level is 0.44~\mjybm. The peak is 11.8~\mjybm  \textit{Bottom: 2008 epoch}.  The rms noise level is 0.18~\mjybm. The peak is 7.9~\mjybm. The contours in the three plots are [$-$3$\sqrt{2}\sigma$, 3$(\sqrt{2})^n\sigma$], where $\sigma$ is the rms noise of each image and n = 1,2,3,...,7}
%\label{fig:WF}
%\end{figure*}
\begin{figure*}[htbp]\scriptsize
\centering
\subfloat{\includegraphics[trim=0cm 2.2cm 0cm 4cm,clip=true,angle=-90,scale=0.48,clip=true]{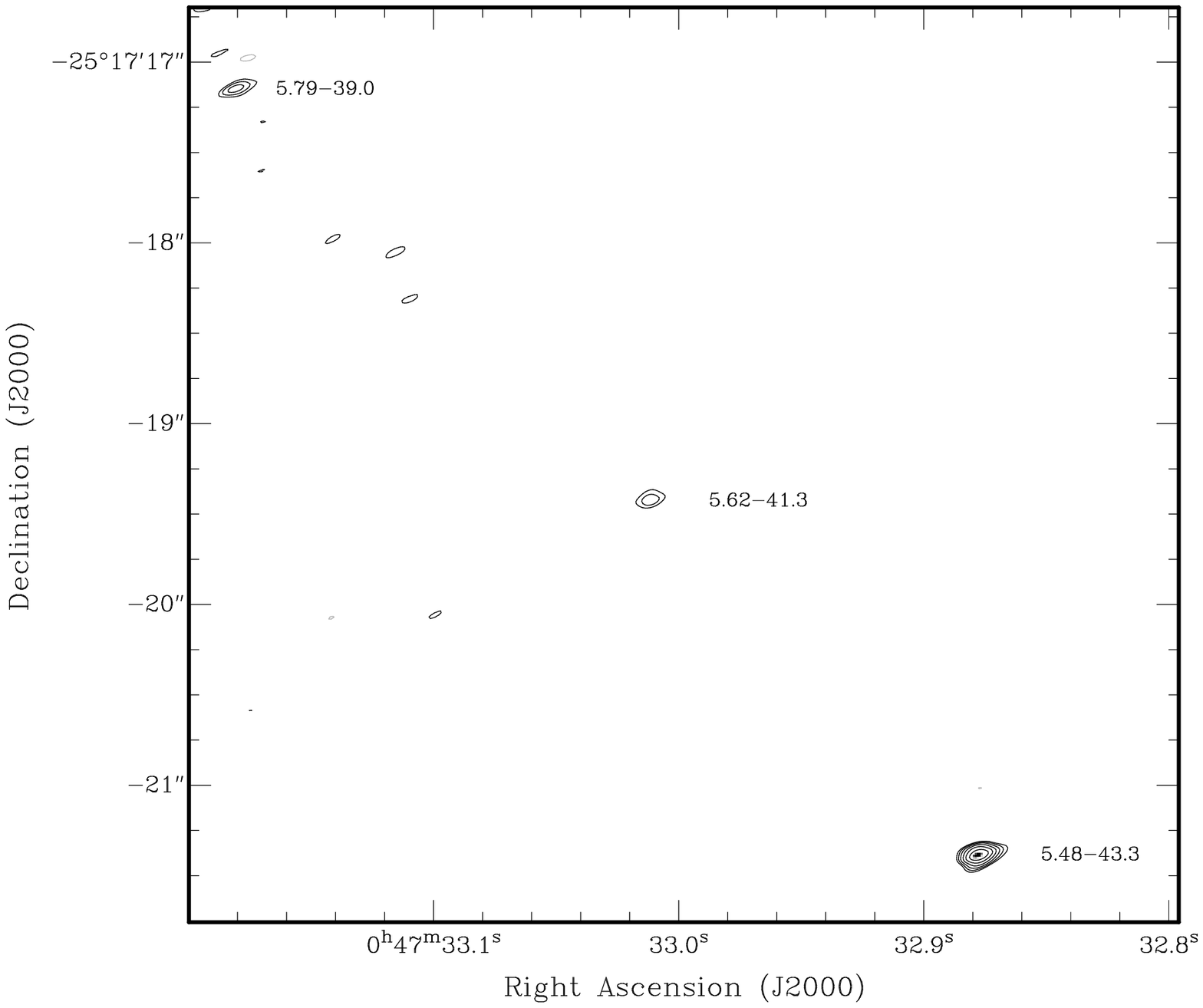}}
\subfloat{\includegraphics[trim=0cm 2.2cm 0cm 0cm,angle=-90,scale=0.48,clip=true]{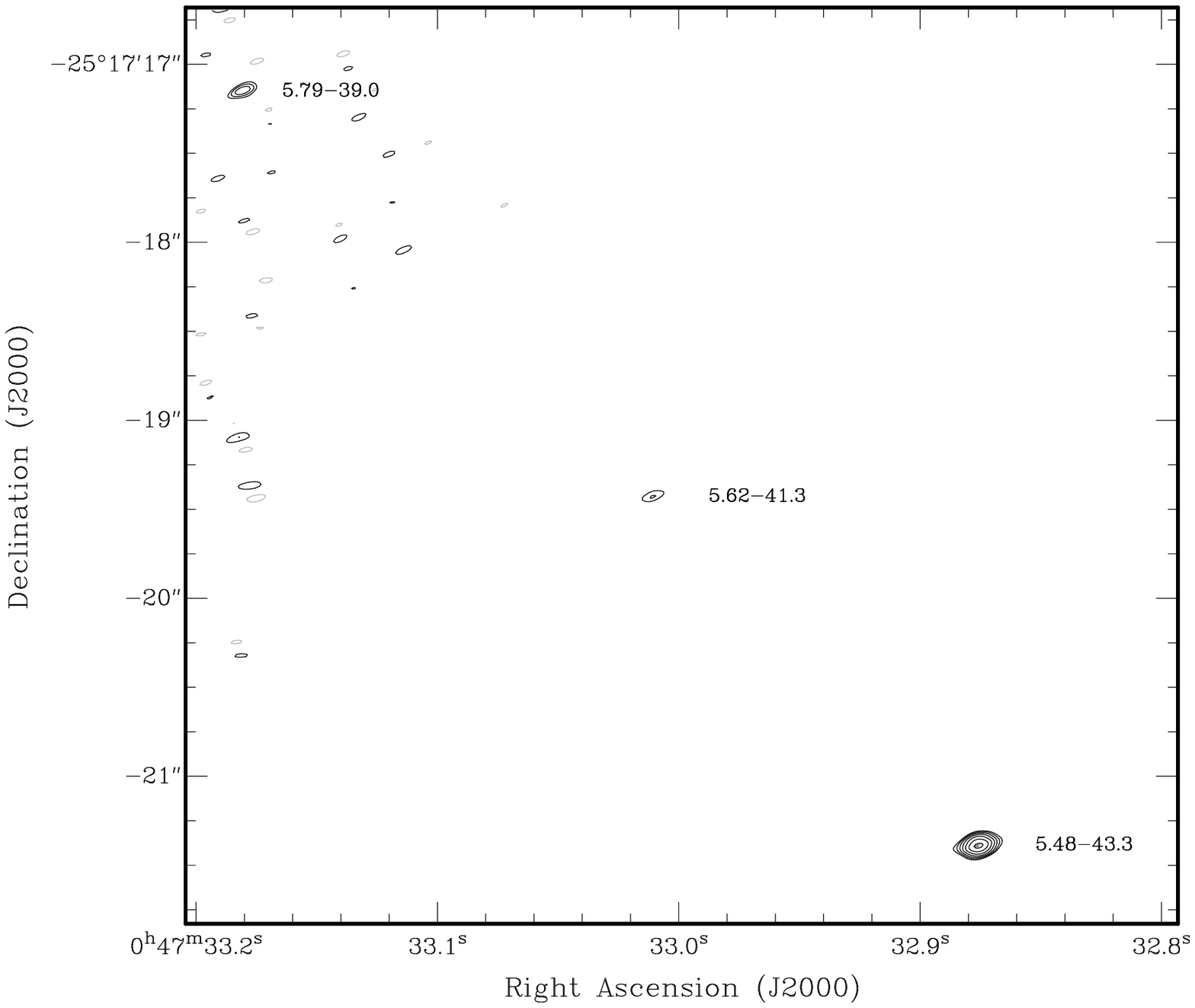}}\\
\subfloat{\includegraphics[trim=0cm 0cm 0cm 0cm,angle=-90,scale=0.48,clip=true]{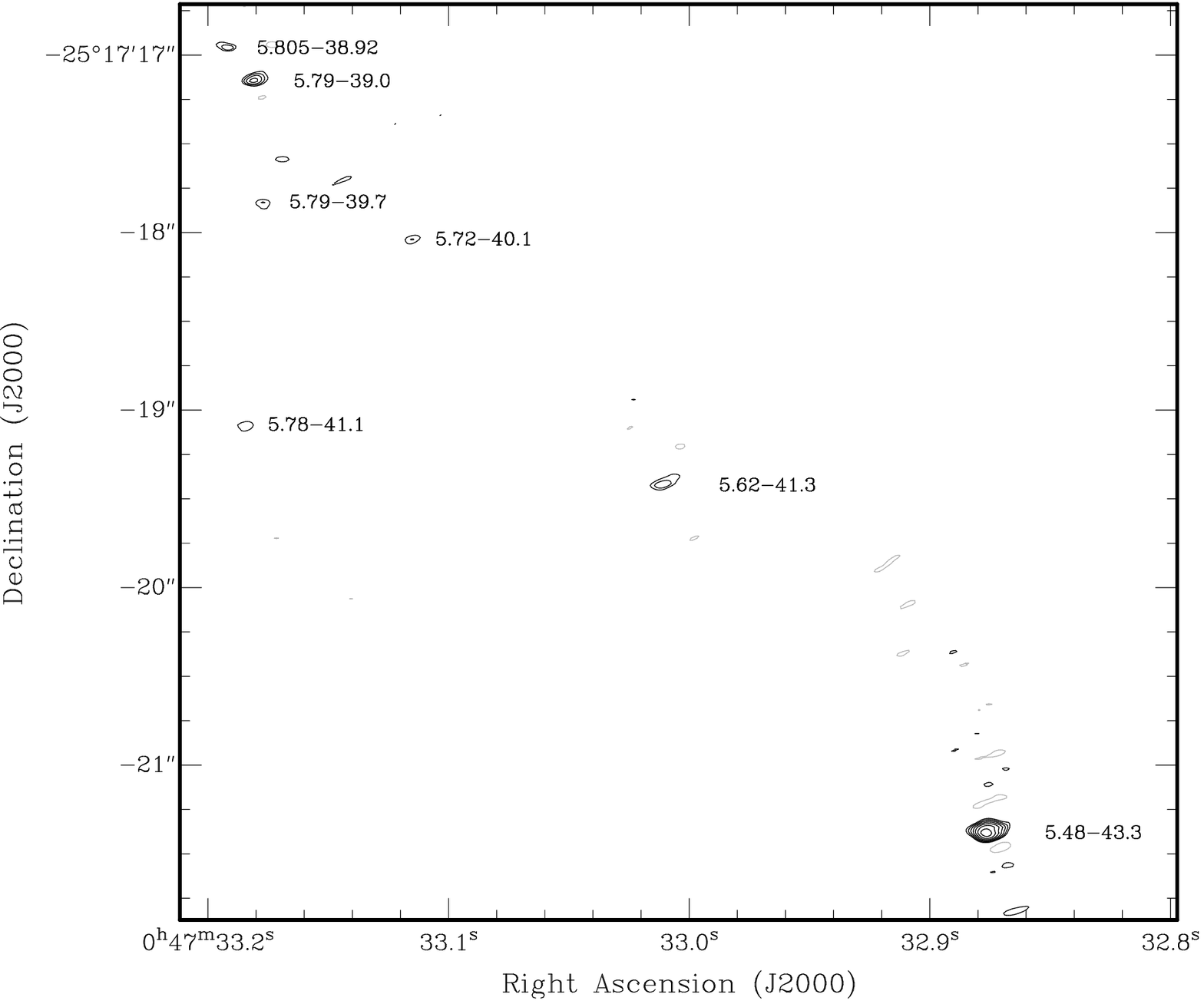}}
\caption{Wide-field VLBI images of NGC~253 with the LBA at 2.3 GHz. \textit{Top Left: 2006 epoch}. The rms noise level is 0.39~\mjybm. The peak is 10.4~\mjybm  \textit{Top Right: 2007 epoch}. The rms noise level is 0.44~\mjybm. The peak is 11.8~\mjybm  \textit{Bottom: 2008 epoch}.  The rms noise level is 0.18~\mjybm. The peak is 7.9~\mjybm. The contours in the three plots are [$-$3$\sqrt{2}\sigma$, 3$(\sqrt{2})^n\sigma$], where $\sigma$ is the rms noise of each image and n = 1,2,3,...,7}
\label{fig:WF}
\end{figure*}
%----------------------------------

%---------------------------------------------------------------------------------------------

%\begin{figure*}[htp]\scriptsize
%\centering
%\begin{tabular}{ccc}
%\includegraphics[trim=0cm 0cm 0cm 0cm,clip=true,angle=-90,scale=0.2]{f4a.eps}&
%\includegraphics[trim=0cm 0cm 0cm 0cm,clip=true,angle=-90,scale=0.2]{f4b.eps}&
%\includegraphics[angle=-90,scale=0.2]{f4c.eps}
% \\
%\end{tabular}
%\caption{2.3~GHz Australian LBA images of 5.48-43.3.  \textit{Left: 2006 epoch.} The rms noise of the image is 0.22~\mjybm and the peak flux density is 1.6~\mjybm. The beam size is 15$\times$14 mas at a position angle (PA) of 0.1\deg. \textit{Centre: 2007 epoch.}  The rms noise of the image is 0.26~\mjybm and the peak flux density is 3.8~\mjybm. The beam size is 15$\times$14 mas at a PA of $-$85\deg. \textit{Right: 2008 epoch.}  The rms noise of the image is 0.13~\mjybm and the peak flux density is 2.61~\mjybm. The beam size is 16$\times$13 mas at a PA of 37\deg. The contours of the three epochs are [$-$3$\sigma_{rms}$, n$\sigma_{rms}$] where, n = 3, 4, 5, 10, 15 and $\sigma_{rms}$ is the respective image rms. }
%\label{fig:5.48}
%\end{figure*}
}

\begin{figure*}[htbp]\scriptsize
\centering
\subfloat{\includegraphics[trim=0cm 0cm 0cm 0cm,clip=true,angle=-90,scale=0.2]{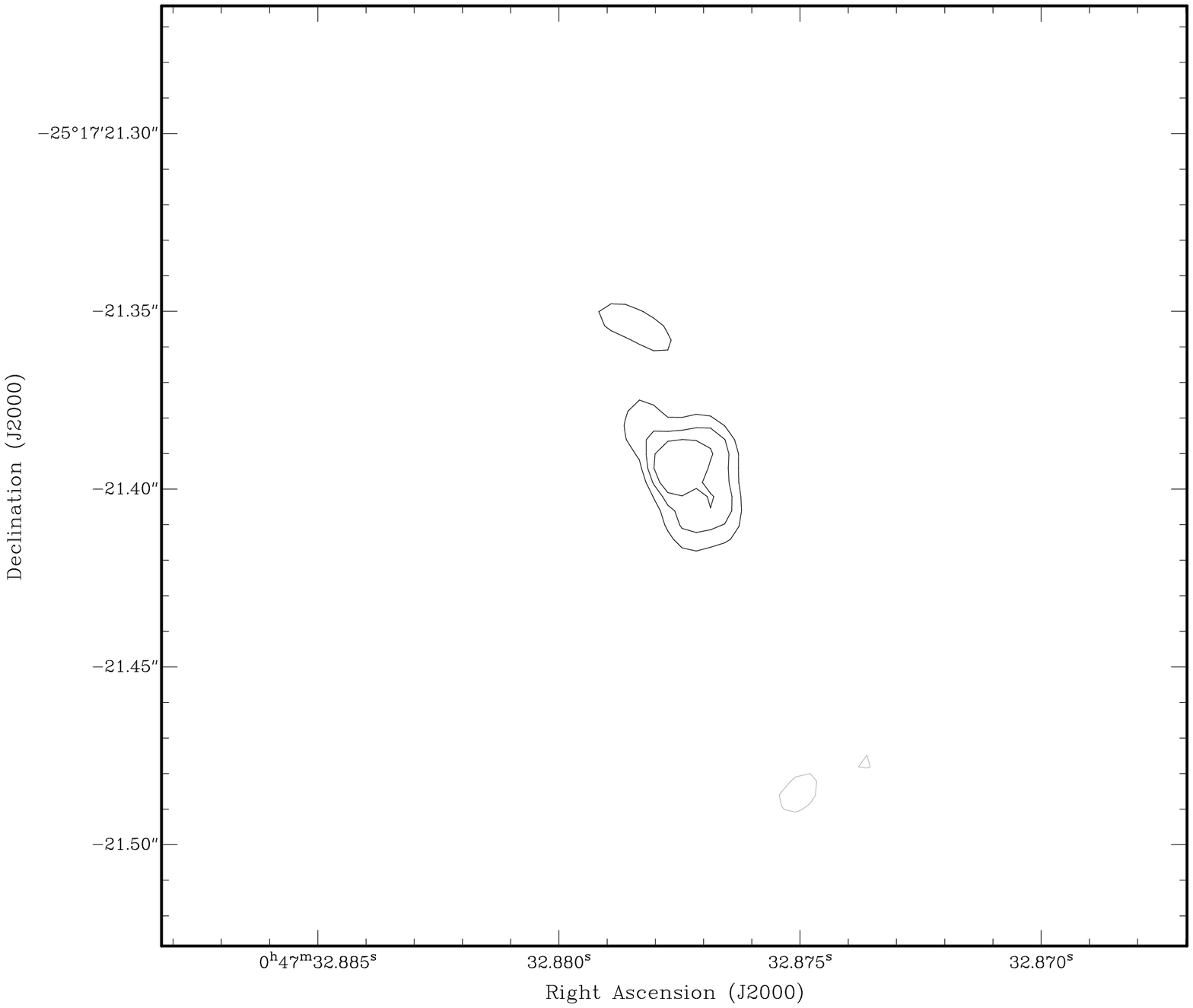}}
\subfloat{\includegraphics[trim=0cm 0cm 0cm 0cm,clip=true,angle=-90,scale=0.2]{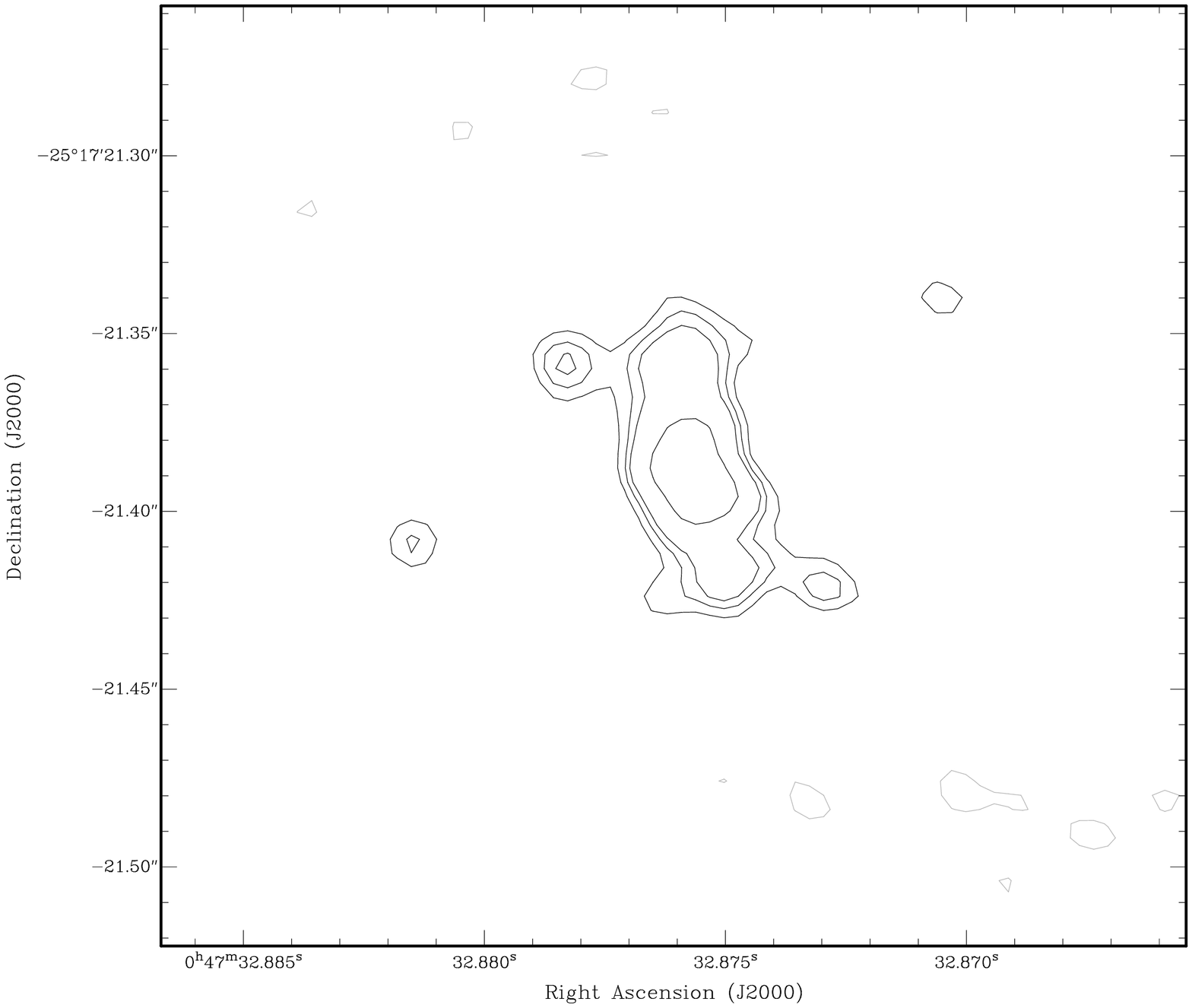}}
\subfloat{\includegraphics[angle=-90,scale=0.2]{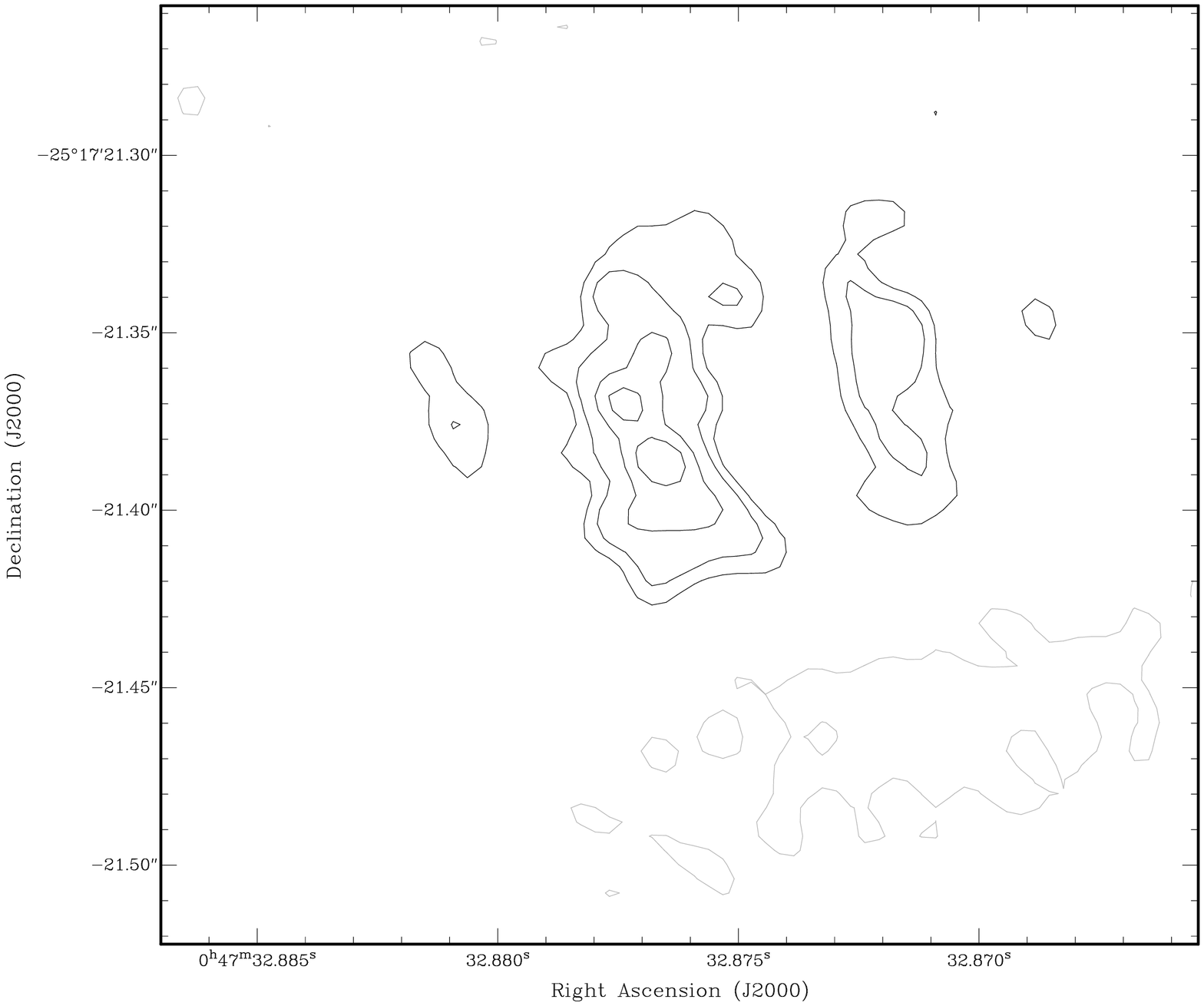}}
\caption{2.3~GHz Australian LBA images of 5.48-43.3.  \textit{Left: 2006 epoch.} The rms noise of the image is 0.22~\mjybm and the peak flux density is 1.6~\mjybm. The beam size is 15$\times$14 mas at a position angle (PA) of 0.1\deg. \textit{Centre: 2007 epoch.}  The rms noise of the image is 0.26~\mjybm and the peak flux density is 3.8~\mjybm. The beam size is 15$\times$14 mas at a PA of $-$85\deg. \textit{Right: 2008 epoch.}  The rms noise of the image is 0.13~\mjybm and the peak flux density is 2.61~\mjybm. The beam size is 16$\times$13 mas at a PA of 37\deg. The contours of the three epochs are [$-$3$\sigma_{rms}$, n$\sigma_{rms}$] where, n = 3, 4, 5, 10, 15 and $\sigma_{rms}$ is the respective image rms. }
\label{fig:5.48}
\end{figure*}

%----------------------------------

The loss of amplitude due to bandwidth smearing at the field's edge (radius of $\sim10^{\prime\prime}$ from the phase centre) is $\sim3\%$ on the longest baseline (Hobart - Ceduna). Thus, to allow imaging of the inner 300~pc at a resolution of $\sim15$~milli-arcseconds (mas), the NGC~253 dataset was not averaged in time or frequency. To facilitate imaging in \textsc{difmap}, the frequency channels were converted into intermediate frequencies (IFs). This conversion allowed \textsc{difmap} to treat the frequency channels independently in the $uv$-plane rather than averaging them together, thus avoiding any further bandwidth smearing effects during the imaging process.

NGC~253 imaging was initially performed with a reduced resolution by excluding data from the Hobart and Ceduna antennas (the longest baselines). The imaging parameters were chosen to closely match the imaging parameters of LT06, with cellsize of 11~mas and application of natural weighting to minimise noise at the expense of resolution. \Fig~\ref{fig:WF} displays the resulting contour images of NGC~253 for the three epochs. The 1$\sigma$ noise measurement for the three images are listed in \Tab~\ref{tab:table1}. The 1$\sigma$ noise levels at the 2006 and 2007 epochs are 2-3 times higher than at the 2008 epoch, which can be attributed to the absence of the 70~m Tidbinbilla antenna in 2006/07. However, the 1$\sigma$ noise of the 2008 image is 28$\%$ lower than that of the LT06 (2004 epoch) image due to increased bandwidth.

\Fig~\ref{fig:5.48} shows the images of the SNR 5.48-43.3 for the three epochs. The images were made with the full array (i.e. including data from both Hobart and Ceduna), with natural weighting and a cell size of 4~mas.

\subsection{Identification of Sources and Flux Density Measurements}
\label{sec:Flux}

\begin{center}
%\begin{savenotes}
\begin{table*}[htb]\scriptsize
%\fbox{% Remove this line to delete the frame
%\begin{minipage}{18cm}
%\renewcommand\footnoterule{}
%{\small
\caption{\textsc{Compact Sources Detected with the LBA at 2.3~ghz}}
 \label{tab:LBAfluxes}
\medskip
%\begin{tabular}{cccccccccccccc}
{\renewcommand{\tabcolsep}{3pt}
\begin{tabular}{*{15}{c}}
\hline
\hline \noalign{\smallskip}
\multicolumn{2}{c}{Position$^{a}$} &  & &  &  &   &  &  & &  &  & & &   \\
\cline{1-2} 
 $\alpha$ (J2000.0) & $\delta$ (J2000.0)& \multicolumn{3}{c}{Identification$^{b}$} & &\multicolumn{4}{c}{Peak Flux Density  (\mjybm) }& &\multicolumn{4}{c}{Integrated Flux Density  (mJy)} \\
 \cline{3-5}\cline{7-10} \cline{12-15}\noalign{\smallskip}
 (00$^{h}$47$^{m})$ & ( $-$25\deg 17\arcm) & UA97& TH85 & LT06$^{c}$ && $S_{P}^{2004\,d}$ &  $S_{P}^{2006}$ &  $S_{P}^{2007}$ &$S_{P}^{2008}$ &&$S_{I}^{2004\,d}$ & $S_{I}^{2006}$ &$S_{I}^{2007}$     &$S_{I}^{2008}$  \\ [1pt]
\hline\noalign{\smallskip}
 32.876 & 21.383 & 5.48$-$43.3 	& TH9 & Y &&  8.6 $\pm$ 0.9 & 13.5$\pm$ 1.6  & 16.0 $\pm$ 1.9 & 7.8 $\pm$ 01.2 & & 32.0 $\pm$ 3.2 &  22.8 $\pm$ 2.4 &  30.2 $\pm$ 3.4 &  25.1 $\pm$ 9.9 \\ [1pt]
 33.012 & 19.425 & 5.62$-$41.3 	& TH7 & Y && 1.7 $\pm$ 0.2 & 2.7 $\pm$ 0.6  &2.4$\pm$ 0.5   & 1.3 $\pm$ 0.3 & &   8.8 $\pm$ 0.9  &  6.3$\pm$ 0.8   &  5.7 $\pm$ 0.8  &  6.8 $\pm$ 2.7 \\ [1pt]
 33.115 & 18.039 & 5.72$-$40.1 	& TH6 & Y && 1.5 $\pm$ 0.2 & $<$2.0 & $<$2.2 & 1.0 $\pm$0.2 & &  3.8 $\pm$ 0.4  &$<$2.0 & $<$2.2 & 3.0 $\pm$ 1.2 \\ [1pt]
 33.185 & 19.084 & 5.78$-$41.1   & ....   & N && $<$1.2 &$<$2.0 & $<$2.2 & 0.9 $\pm$0.2 && $<$1.2 &$<$2.0 & $<$2.2  & 4.6 $\pm$ 1.8   \\[1pt]
 33.177 & 17.830 & 5.79$-$39.7 	& TH3 & Y && 1.8 $\pm$ 0.2 & $<$2.0 & $<$2.2 & 1.0 $\pm$0.2 & & 6.2 $\pm$ 0.6  & $<$2.0 & $<$2.2 & 3.2 $\pm$ 1.3 \\ [1pt]
 33.181 & 17.148 & 5.79$-$39.0 	& TH2 & Y && 2.4 $\pm$ 0.2 & 4.1 $\pm$ 0.6  & 4.9 $\pm$ 0.7  &2.6 $\pm$0.4 && 5.7 $\pm$ 0.6  &  7.8 $\pm$ 1.0  &  7.3 $\pm$ 0.9  &   5.9 $\pm$ 2.3\\ [1pt]
% 33.193 & 16.949 & 5.805$-$38.92 & ... & Y && 2.1 $\pm$ 0.2 & $<$2.0 & $<$2.2 & 1.5 $\pm$ 0.3 && 6.8 $\pm$ 0.7  &$<$2.0 & $<$2.2 & 2.4 $\pm$ 1.0\\ 
 \hline
\\
\multicolumn{15}{l}{$^{a}$The measured 2.3 GHz source positions for the 2008 epoch. Units of right ascension are seconds and units of declination are arcseconds.}\\
\multicolumn{15}{l}{$^{b}$Source identifications with 1.3 and 2 cm sources from UA97, and 2~cm sources from TH85.}\\
\multicolumn{15}{l}{$^{c}$Detected (Y) or not-detected (N) by LT06.}\\
\multicolumn{15}{l}{$^{d}$Acquired from LT06.}\\

\end{tabular}
}
\hfill{}

 %\end{minipage}%
 % }
\end{table*}
%\end{savenotes}
\end{center}
%\end{landscape}
%\pagebreak

%----------------------------------------------------------------------------------------------------------------------------------------------------------------------------------

A number of detected sources are clearly visible in \fig~\ref{fig:WF}. We used the source extraction software, \textsc{blobcat} \citep{blobcatpaper} to identify and measure the flux densities of compact sources in the three images. Seven sources were detected above 5$\sigma$ by \textsc{blobcat} in the 2008 epoch, while only the three brightest sources (5.48-43.3, 5.62-41.3 and 5.79-39.0) were found via \textsc{blobcat} for the less sensitive 2006 and 2007 epochs (see \Tab~\ref{tab:LBAfluxes}).

Included in \Tab~\ref{tab:LBAfluxes} are the flux densities for the compact sources detected in the 2004 epoch by LT06.  Only three of these sources were detected in all four 2.3~GHz LBA epochs (i.e. 2004, 2006, 2007 and 2008), while three were detected in only the 2004 and 2008 epochs, with one possible additional source detected solely in the 2008 image. The flux density of the sources detected by LT06 were also estimated via \textsc{blobcat} and found to agree with LT06's values within measurement errors.  For comparison, the published source flux densities of LT06 are included in \Tab~\ref{tab:LBAfluxes}. While stacking of the 2004 and 2008 images improved image noise by a factor of $\sqrt{2}$, the resulting image did not reveal any additional sources.

Errors in the flux-density estimates have two different origins. Firstly, there is the overall uncertainty in the absolute flux density scale for the LBA, estimated as  $\pm\,10\%$  \citep{Reynolds1994}. Secondly, there are the errors in determining the flux density by \textsc{blobcat}, which ranges from 5 - 10$\%$. Both terms are added in quadrature for the 2006 and 2007 observations. For the 2008 observation, we determine a third contribution to the error, which is described in \Sect~\ref{sec:fluxvar}.

Table~\ref{tab:LBAfluxes} lists cross identifications with sources detected at 15~GHz and 23~GHz by UA97 and at 2.3~GHz with the LBA by LT06. The mean position difference between our 2008  and LT06's 2004 J2000.0 source positions was found to be 8~mas, with a  standard deviation of 5~mas, which is $\sim10\%$ of the beam-width. The mean position difference between our 2008 J2000.0 source position and the B1950.0 source positions of the 23 GHz sources from UA97 is 87~mas, with a standard deviation of 51~mas.

\placetable{tab:LBAfluxes}
\subsubsection{Variations in the Flux Density}
\label{sec:fluxvar}

%------------------------------------------------------------------------------
\begin{figure}[htp]
\centering
\includegraphics[trim=1cm 0cm 0cm 0cm,clip=true,scale=0.45]{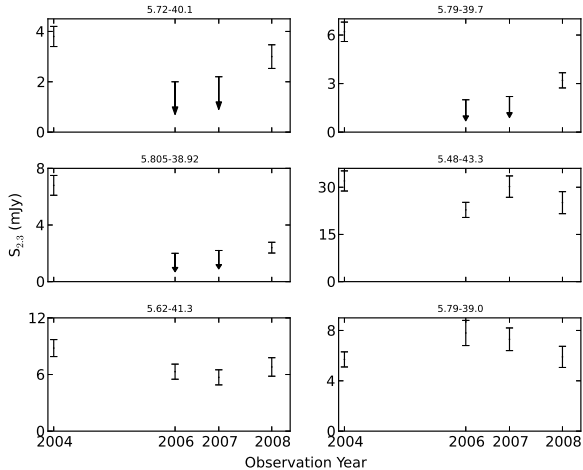} 
\caption{Light curves of the compact sources detected in all 2.3 GHz observations (2004, 2006, 2007 and 2008). The integrated flux densities (from \Tab~\ref{tab:LBAfluxes}) are plotted on the y-axis, while the observation date is listed on the x-axis.  5$\sigma$ upper limits are represented by arrows for epochs without detections. The error-bars are the absolute flux density scaling and measurement errors (see \Sect~\ref{sec:Flux}) added in quadrature.}
\label{fig:lcurve}
\end{figure}
%------------------------------------------------------------------

Lower integrated flux densities are recorded for all sources except for 5.79$-$39.0 in the 2008 epoch compared to the 2004 epoch (see \Tab~\ref{tab:LBAfluxes} and \Fig~\ref{fig:lcurve}). Since it was necessary to interpolate the flux density of a variable calibrator to obtain a value for the 2008 epoch (see \Sect~\ref{sec:Calib}) it would not be surprising if there were an error on the flux density scale. Given the covariance of the flux densities between 2006 and 2008 (and possibly for the other epochs too) this would appear to be the case. Adjusting empirically for the difference in flux density between 2006 and 2008 by taking the mean of each, we find that we must increase the absolute flux density calibration of the 2008 epoch by 20$\%$. After this adjustment, no source appears significantly variable with the exception of 5.805-38.92, whose decrease in flux density would still be a 9$\sigma$ error (assuming of course that the distribution of the errors is Gaussian).

\begin{figure}[htp]
\centering
\includegraphics[trim=0cm 2.5cm 0cm 3.5cm,clip=true,scale=0.42]{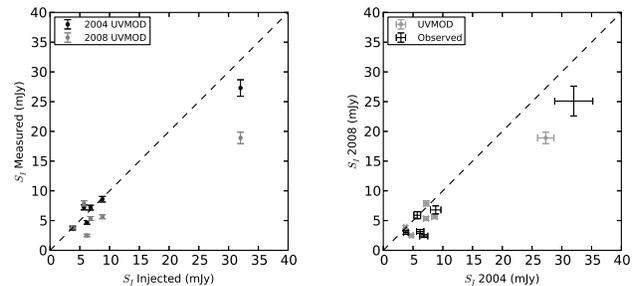} 
\caption{\textit{Left:} The integrated flux densities of the fake sources (x-axis) injected into the modelled datasets via \textsc{uvmod} are plotted against the recovered integrated flux densities for the 2004 (black points) and 2008 (grey points) modelled datasets. \textit{Right:} The recovered integrated flux densities of the fake sources for the 2004 and 2008 modelled datasets (grey points) are compared, while the observed integrated flux densities from \Tab~\ref{tab:LBAfluxes} for the same epochs (black points). The dashed lines represents where y=x. }
\label{fig:uvmod}
\end{figure}

To investigate whether the flux density decrease is genuine we used the \textsc{aips} task \textsc{uvmod} to simulate datasets with the same $uv$-coverage and noise level as the 2004 and 2008 LBA datasets. In \textsc{uvmod} the parameter \textsc{factor} was set to zero to allow the injection of the fake sources at the same positions as the real ones. 
The fake sources were injected with the properties (i.e. size, position angle, integrated flux density and position) of the sources detected by LT06. The resulting datasets were exported and imaged in \textsc{difmap} following the procedure described in \Sect~\ref{sec:Calib}.  
%To estimate the noise, the parameter \textsc{flux} was set to the sensitivity of the least sensitive baseline (ATCA-MP = 3.5~mJy and 2.5 mJy, for the 2004 and 2008 epochs).
The integrated flux densities of the injected sources were measured with \textsc{blobcat} and plotted in \Fig~\ref{fig:uvmod} (left plot) against the model flux densities. If the flux density decrease in the observed dataset were due to only source variability, the measured (or recovered) flux densities would be equal to the injected flux densities, represented by the dashed lines. The deviation of the \textsc{uvmod} flux densities from the dashed lines follows a similar trend to the observed flux densities (right plot of \Fig~\ref{fig:uvmod}). Thus it is possible the flux density decrease results mainly from systematic effects such as deconvolution errors (e.g. \textsc{clean} bias \citealt{Beckeretal95}) and high side-lobes.

%------------------------------------------------------------------------------------------------

%---------------------------------------

The effect of \textsc{clean} bias is known to produce a systematic underestimate of flux densities \citep{Beckeretal95} by redistributing flux density from sources to noise peaks in the image during deconvolution. The magnitude of the flux density redistributed is generally independent of flux density, and thus the fractional error is largest for weak sources. This effect may have a stronger effect in the 2008 image (as compared to the 2004 image) due to the presence of high side-lobes. The high side-lobes may have resulted from the combination of poor $uv$-coverage and an increased sensitivity due to a higher bandwidth.
 
The RMS\footnote{We define the root mean square (RMS) difference between the modelled and the recovered flux densities, as $\sqrt{1/n\Sigma(S_{model}-S_{recovered})/S_{model}}$} difference between the 2008 recovered flux densities and the model flux densities is 37$\%$, which is added in quadrature with the absolute flux density scaling and measurement errors for the 2008 epoch (see \Sect~\ref{sec:Flux}) and listed in \Tab~\ref{tab:LBAfluxes}. The resulting flux density decrease of 5.805$-$38.92 is then reduced to a significance of 4.5$\sigma$. While this variability may be due to unaccounted for calibration errors, we cannot rule out contributions from intrinsic variability.

%-------------------------------------------------------------------
%\begin{landscape}
\begin{center}
%\begin{savenotes}
\begin{table*}[htp]\scriptsize
\caption{\textsc{Summary of Flux Density Measurements for Radio Sources in ngc~253}}
\label{tab:allfluxes}
\medskip
\hfill{}
{\renewcommand{\tabcolsep}{4pt}
\begin{tabular}{ccccccccccccc}
\hline
\hline
\noalign{\smallskip}
& \multicolumn{5}{c}{LBA$^{a}$ (mJy)} 
&\multicolumn{5}{c}{VLA$^{b}$ (mJy)}
 \smallskip\\
\cline{2-6}
\cline{8-12}
\noalign{\smallskip}
 VLA ID &  $S_{1.4}$ & $S_{2.3}^{2004}$ & $S_{2.3}^{2006}$ & $S_{2.3}^{2007}$ & $S_{2.3}^{2008}$ &&$S_{5}$ & $S_{8.3}$ & $S_{15}$ &$S_{23}$ & $S_{23}^{BCT}$  \smallskip \\ 
\hline

4.81-43.6 & $<$1.8 & $<$1.2 & $<$2.0 & $<$2.2  & $<$0.9 && 1.4 $\pm$ 0.2 & ....    & 0.4 $\pm$ 0.3   & ....  &  1.4 $\pm$ 0.3\\ 
5.48-43.3 & 16 $\pm$ 0.2        & 32.0 $\pm$ 3.2     & 22.8 $\pm$ 2.4     & 30.2  $\pm$ 3.4     & 25.1  $\pm$ 9.9    &&27.1 $\pm$ 2.7 & 20.5 $\pm$  1.0 & 12.5 $\pm$ 1.3 & 9.8 $\pm$ 1.0 & 13.1 $\pm$ 0.9\\ 
5.49-42.2 & $<$1.8 & $<$1.2 & $<$2.0 &$<$2.2   & $<$0.9 && ....    & 1.1 $\pm$ 0.2  & ....     & 0.8 & ....    \\ 
5.54-42.2 & $<$1.8 & $<$1.2 & $<$2.0 & $<$2.2& $<$0.9 && 2.8 $\pm$ 0.4 & 2.7 $\pm$  0.24 & 3.4 $\pm$ 0.4 & 4.5 $\pm$ 0.5 & 9.2 $\pm$ 1.7 \\ 
5.59-41.6 & $<$1.8 & $<$1.2 & $<$2.0 & $<$2.2 & $<$0.9 && .... & 3.0 $\pm$  0.3 & .... & 5.7 & .... \\ 
5.62-41.3 & 6 .0 $\pm$ 0.6        & 8.8 $\pm$ 0.9     & 6.3 $\pm$ 0.8 & 5.7 $\pm$ 0.8 & 6.8 $\pm$ 2.7 && 9.8 $\pm$ 1.0 & 7.3 $\pm$  0.4 & 7.5 $\pm$ 0.8 & 5.8 $\pm$ 0.6 & 10.7 $\pm$ 1.4 \\ 
5.65-40.7 & $<$1.8 & $<$1.2 & $<$2.0 & $<$2.2& $<$0.9 && .... & 1.5 $\pm$  0.2 & .... & 1.2  $\pm$ 0.3 & .... \\ 
5.72-40.1 & $<$1.8 & 3.8 $\pm$ 0.4     & $<$2.0 & $<$2.2 & 3.0 $\pm$ 1.2 && 7.7 $\pm$ 0.8  & 6.5 $\pm$  0.4 & 7.8 $\pm$ 0.8 & 7.9 $\pm$ 0.8 & 21.9 $\pm$ 2.3 \\ 
5.73-39.5 & $<$1.8 & $<$1.2 & $<$2.0 & $<$2.2 & $<$0.9 && .... & 2.4 $\pm$  0.2 & .... & 2.8 $\pm$ 0.4 & ....\\ 
5.75-41.8 & $<$1.8 & $<$1.2 & $<$2.0 & $<$2.2 & $<$0.9 && 7.0 $\pm$ 0.7 & 4.8 $\pm$  0.3 & 3.1 $\pm$ 0.4 & 2.6 $\pm$ 0.4 & .... \\ 
5.78-39.4 & $<$1.8 & $<$1.2 & $<$2.0 & $<$2.2 & $<$0.9 && .... & 16.9 $\pm$  0.9 & .... & 9.7 $\pm$ 1.0 &  29.9$\pm$ 1.2 \\ 
5.79-39.0 & $<$1.8 & 5.7 $\pm$ 0.6 & 7.8 $\pm$ 1.0 & 7.3 $\pm$ 0.9 & 5.9 $\pm$ 2.3 && 38.6 $\pm$  3.9 & 48.0 $\pm$ 2.4 & 40.3 $\pm$ 4.0 &35.8 $\pm$ 3.6 & 38.4 $\pm$ 0.9 \\ 
5.79-39.7 & $<$1.8 & 6.2 $\pm$ 0.6 & $<$2.0 & $<$2.2 & 3.2 $\pm$ 1.3 & &.... & .... & $\sim$4.1 $\pm$ 0.5 & $\sim$1.6 $\pm$ 0.3 & .... \\ 
5.78-41.1 & $<$1.8 & $<$1.2 & $<$2.0 & $<$2.2 & 4.6 $\pm$ 1.8 && 4.7 $\pm$  0.5 & 3.2 $\pm$ 0.3 & 1.9 $\pm$ 0.3 & 1.0 $\pm$ 0.3  & .... \\ 
5.805-38.92 & $<$1.8 & 6.8  $\pm$ 0.7   & $<$2.0 & $<$2.2 & 2.4 $\pm$ 1.0 && .... & .... & $\sim$3.0 $\pm$ 0.4 & $\sim$1.9  $\pm$ 0.3 &  8.9$\pm$ 1.1\\ 
5.87-40.1 & $<$1.8 & $<$1.2 & $<$2.0 & $<$2.2 & $<$0.9 && 3.8 $\pm$ 0.4 & 2.1 $\pm$ 0.2 & 1.4 $\pm$ 0.3 & 0.7 $\pm$ 0.3 & .... \\ 
5.90-37.4 & $<$1.8 & $<$1.2 & $<$2.0 & $<$2.2 & $<$0.9 && 4.0 $\pm$ 0.5 & 4.0 $\pm$ 0.3 & 5.9 $\pm$ 0.7 & 6.7 $\pm$ 0.7 &  9.1 $\pm$ 1.2 \\ 
5.95-37.7 & $<$1.8 & $<$1.2 & $<$2.0 & $<$2.2 & $<$0.9 && 1.2 $\pm$ 0.2 & .... & 0.6 $\pm$ 0.3 & .... & .... \\ 
6.00-37.0 & $<$1.8 & $<$1.2 & $<$2.0 & $<$2.2 & $<$0.9 && 5.9 $\pm$ 0.6 & 3.5 $\pm$ 0.2 & 2.4 $\pm$ 0.4 & 1.7 $\pm$ 0.3& ....\\ 
6.40-37.1 & $<$1.8 & $<$1.2 & $<$2.0 & $<$2.2 & $<$0.9& & 2.9 $\pm$ 0.4& .... & 1.3 $\pm$ 0.3 & .... &  0.8$\pm$0.2 \\ 

\hline
\\
\multicolumn{13}{l}{$^{a}S_{1.4}$ is the measured flux density from the LBA \citep{Tingay04}, $S_{2.3}$ are the measured flux density at 2.3~GHz from 2004 (LT06) and this}\\
\multicolumn{13}{l}{work (2006-2008), where the superscript indicates the observation year. Upper limits on the flux (5$\sigma$) are provided for sources that have } \\
\multicolumn{13}{l}{not been detected.}\\
\multicolumn{13}{l}{$^{b}S_{5}$, $S_{8.3}$,  $S_{15}$ and $S_{23}$ are the measured 5, 8.3, 15 and 23~GHz flux densities from the VLA by UA97 and $S_{23}^{BCT}$  are the 23~GHz VLA flux}\\
\multicolumn{13}{l}{densities of \citet{Brunthaleretal2009}.}\\
\end{tabular}
}
\hfill{}
\end{table*}
\end{center}

%------------------------------------------------

\begin{figure*}[htp]
\centering
\includegraphics[scale=0.9]{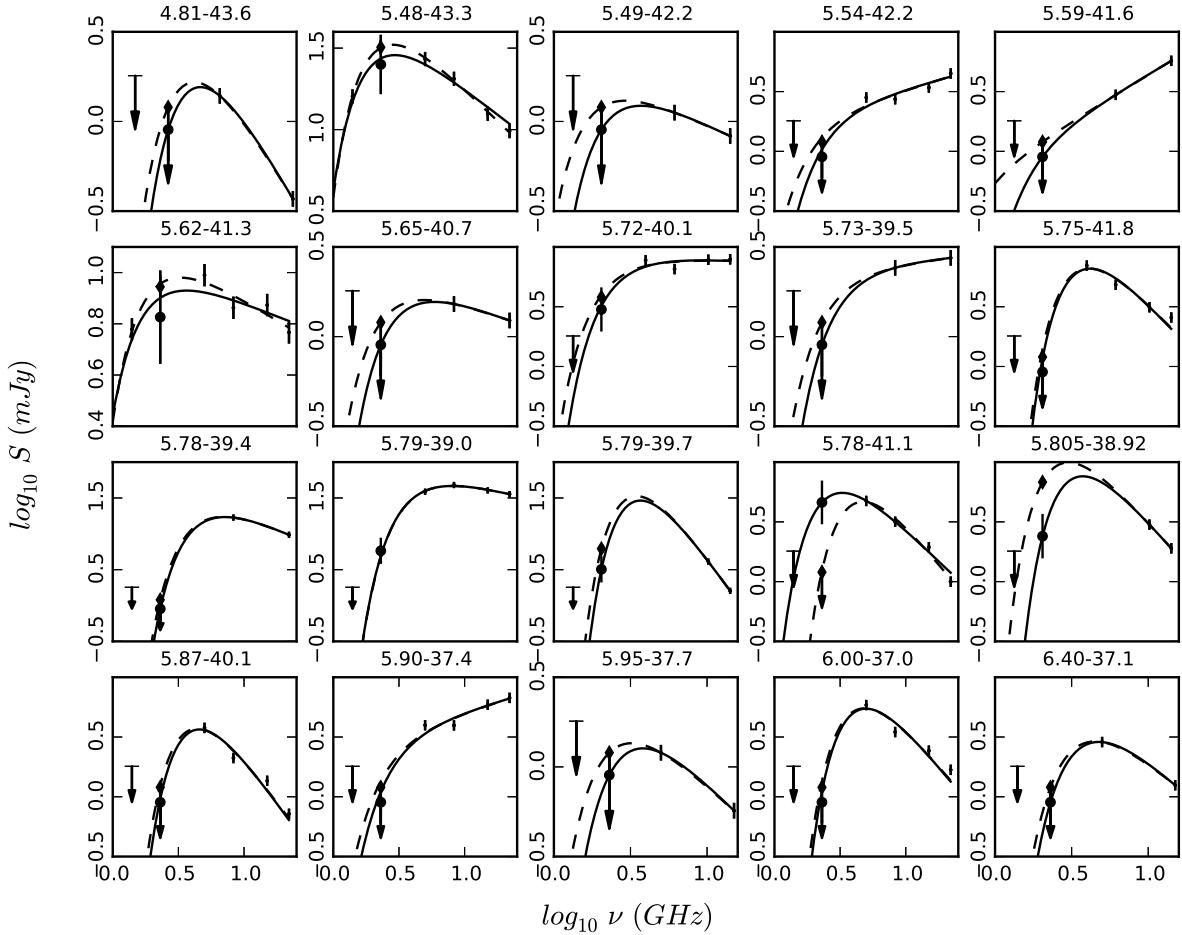}
\caption{Measured flux densities (symbols with error bars) and free-free absorption models (lines) for 20 UA97 sources. The solid lines plots the free-free model derived using the 2008 2.3 GHz flux density values (star points). For comparison the free-free model derived with LT06's 2004 2.3~GHz flux density values (dashed lines) are included. The 3 and 5 $\sigma$ upper limits are shown for sources not detected at 1.4~GHz (1.8~mJy) and 2.3~GHz (1.2~mJy [2004] and 0.9~mJy [2008]) }
 \label{fig:ffspectra}
\end{figure*}

\begin{center}
\begin{table*}[ht]\scriptsize
\begin{minipage}{18cm}
\renewcommand\footnoterule{}
\caption{\textsc{Parameters of the free-free absorption models for all compact sources.}}
\label{tab:ffparams}
\medskip
\hfill{}
\begin{tabular}{l|cccc|cccc}
\hline\hline
\noalign{\smallskip}
Source & $S_{0}^{2004}$ & $\alpha^{2004}$ & $\tau^{2004}$ & Type$^{a}$ & $S_{0}^{2008}$ &  $\alpha^{2008}$& $\tau^{2008}$ & Type 
\smallskip\\ \hline
4.81-43.6 & 19.92$\pm$12.81 &$-$1.46$\pm$0.62 & 9.16$\pm$5.16 & S & 24.1$\pm$16.4 & $-$1.53$\pm$0.73 & 11.59$\pm$6.84 & S \\ 
5.48-43.3 & 105.6$\pm$0.3 & $-$0.77$\pm$0.01 & 3.30$\pm$0.02 & S & 76.4$\pm$0.13 & $-$0.63$\pm$0.01 & 2.89$\pm$0.01 & S \\ 
5.49-42.2 & 2.31$\pm$0.53 & $-$0.33$\pm$0.15 & 2.19$\pm$1.65 & T & 2.47$\pm$0.84 & $-$0.35$\pm$0.24 & 4.12$\pm$3.11 & T \\ 
5.54-42.2 & 1.54$\pm$0.18 & 0.33$\pm$0.06 & 2.54$\pm$1.66 & T & 1.68$\pm$0.30 & 0.3$\pm$0.09 & 4.21$\pm$2.60 & T \\ 
5.59-41.6 & 0.72$\pm$0.04 & 0.67$\pm$0.03 & 0.29$\pm$1.12 & T & 0.77$\pm$0.11 & 0.65$\pm$0.07 & 2.23$\pm$2.30 & T \\ 
5.62-41.3 & 16.11$\pm$0.04 & $-$0.31$\pm$0.01 & 1.82$\pm$0.02 & T & 12.25$\pm$0.07 & $-$0.21$\pm$0.01 & 1.49$\pm$0.02 & T \\ 
5.65-40.7 & 2.71$\pm$0.73 & $-$0.25$\pm$0.17 & 3.46$\pm$2.28 & T & 2.89$\pm$1.02 & $-$0.27$\pm$0.24 & 5.40$\pm$3.66 & T \\ 
5.72-40.1 & 8.25$\pm$0.37 & $-$0.02$\pm$0.02 & 4.12$\pm$0.53 & T & 9.03$\pm$0.94 & $-$0.05$\pm$0.04 & 5.70$\pm$1.23 & T \\ 
5.73-39.5 & 2.07$\pm$0.47 & 0.09$\pm$0.12 & 3.59$\pm$2.33 & T & 2.21$\pm$0.65 & 0.07$\pm$0.17 & 5.52$\pm$3.58 & T \\ 
5.75-41.8 & 76.97$\pm$5.12 & $-$1.16$\pm$0.04 & 17.46$\pm$0.64 & S & 86.16$\pm$5.80 & $-$1.20$\pm$0.04 & 18.98$\pm$0.66 & S \\ 
5.78-39.4 & 121.18$\pm$2.49 & $-$0.81$\pm$0.01 & 22.67$\pm$0.51 & S & 129.41$\pm$3.35 & $-$0.83$\pm$0.01 & 24.61$\pm$0.67 & S \\ 
5.79-39.0 & 150.79$\pm$0.59 & $-$0.46$\pm$0.01 & 17.58$\pm$0.08 & S & 149.60$\pm$2.32 & $-$0.46$\pm$0.01 & 17.45$\pm$0.26 & S \\ 
5.79-39.7 & 4899.0$\pm$2737.6 & $-$2.58$\pm$0.93 & 25.15$\pm$10.52 & S & 5244.7$\pm$2959.9 & $-$2.6$\pm$1.14 & 30.09$\pm$13.9 & S \\ 
5.78-41.1 & 75.97$\pm$5.53 & $-$1.38$\pm$0.04 & 17.09$\pm$0.60 & S & 33.03$\pm$4.71 & $-$1.07$\pm$0.08 & 6.20$\pm$0.88 & S \\ 
5.805-38.92 & 87.74$\pm$25.72 & $-$1.23$\pm$0.15 & 8.26$\pm$1.68 & S & 97.7$\pm$49.61 & $-$1.27$\pm$0.55 & 15.24$\pm$7.14 & S \\ 
5.87-40.1 & 70.27$\pm$36.82 & $-$1.50$\pm$0.51 & 16.00$\pm$6.63 & S & 84.0$\pm$16.12 & $-$1.57$\pm$0.11 & 18.17$\pm$1.33 & S \\ 
5.90-37.4 & 2.30$\pm$0.35 & 0.35$\pm$0.07 & 4.72$\pm$2.30 & T & 2.48$\pm$0.48 & 0.32$\pm$0.09 & 6.20$\pm$3.05 & T \\ 
5.95-37.7 & 5.08$\pm$1.94 & $-$0.80$\pm$0.30 & 4.45$\pm$2.67 & S & 6.15$\pm$2.89 & $-$0.87$\pm$0.42 & 6.89$\pm$4.37 & S \\ 
6.00-37.0 & 82.41$\pm$8.22 & $-$1.32$\pm$0.05 & 17.28$\pm$0.80 & S & 94.1$\pm$9.4 & $-$1.37$\pm$0.05 & 18.98$\pm$0.82 & S \\ 
6.40-37.1 & 25.72$\pm$12.40 & $-$1.10$\pm$0.47 & 12.35$\pm$5.91 & S & 31.17$\pm$15.94 & $-$1.17$\pm$0.55 & 14.78$\pm$7.40 & S \\ 
\hline\\
\multicolumn{9}{l}{The superscript indicates the LBA 2.3~GH epoch.}\\
\multicolumn{9}{l}{$^{a}$LT06 classification of radio sources in NGC 253: S=Supernova Remnant; T=Thermally dominated H~\textsc{ii} region.}\\
\end{tabular}
\hfill{}
 \vspace{-2ex}   
 \end{minipage}%
\end{table*}
\end{center}

\section{Discussion}

\subsection{Radio Spectra and Free-Free Absorption Modelling of Compact Sources in NGC253}
\label{sec:CS}

A downturn in the spectra of the compact sources in NGC~253 at low frequencies was observed previously by \citet{Tingay04} and LT06. Different mechanisms were explored by LT06 to explain the observed effect including; a simple power law, a power law spectrum with free-free absorption by a screen of ionized gas, and a self-absorbed bremsstrahlung spectrum. LT06 demonstrated that the spectra of the sources were consistent with a free-free absorbed power-law spectrum, given by the following equation: 

\begin{equation}
S(\nu) = S_{0}\nu^{\alpha}e^{-\tau(\nu)},
\label{eq:spectra2}
\end{equation}
where
\begin{equation}
\tau(\nu) = \tau_{0}\nu^{-2.1}
\label{eq:tau}
\end{equation}

and $\alpha$ is the optically thin intrinsic spectral index, $\tau_{0}$ is the free-free optical depth at 1~GHz, and $S_{0}$ is the intrinsic flux density of the source at 1~GHz. Similar spectra have been obtained for compact sources in other nearby starburst galaxies, such as M82 \citep{Mcdonaldetal02}, Arp~220 \citep{Paraetal2007}, and NGC~4945 \citep{LT09}. We therefore adopt this model and the method described by LT06.

To model the free-free absorption towards the compact sources within NGC~253 we compile multi-wavelength radio flux density measurements for these sources from the literature (listed in \Tab~\ref{tab:allfluxes}), including: 1.4~GHz LBA data \citep{Tingay04}; 2.3~GHz LBA data (LT06 and this paper); 23~GHz, 15 GHz, 8.3~GHz and 5~GHz VLA data (UA97).

\citet{Brunthaleretal2009} detected 10 UA97 sources at 23~GHz with the VLA (A configuration) in 2004. They reported higher flux densities for all sources compared to UA97, at the same frequency. Only the source 5.79$-$39.0 (the assumed core TH2) was similar in both epochs within a 10$\%$ error. The authors attributed the higher flux densities in their data to a larger beam ($\sim 40\%$). To keep the spectral modelling consistent with LT06, and maintain a comparable beam size with the LBA 2.3 GHz images, the flux densities from \citet{Brunthaleretal2009} were not used.

\Fig~\ref{fig:ffspectra} presents multi-epoch monitoring of the free-free spectra of 20 compact sources in NGC~253. The free-free spectra obtained by LT06 are plotted as dashed lines, while the solid lines are the spectra derived using the 2008 2.3~GHz LBA data. For sources not detected in the 2008 observation, an upper limit of 0.9~\mjybm ~(5$\sigma$) was adopted. Table~\ref{tab:ffparams} lists the fitted free parameters $\alpha$, $\tau_{0}$, and $S_{0}$ for each compact source. Included for reference are the values of the parameters for the 2004 data. Differences in the free-free spectra and derived parameters between 2004 and 2008, while potentially large, are still within our errors, suggesting no evidence for changes in the free-free absorbing medium between epochs. These results further confirm the free-free absorption interpretation of \citet{Tingay04} and LT06. 

From the radio spectra, LT06 deduced that eight\footnote{A typographical error was made in table 3 of LT06. The spectral index of 5.805-38.92 should be -1.20 and not 1.20} of the 20 sources were thermally dominated H~\textsc{ii} regions (indicated with a flat intrinsic spectrum, $\alpha\,>~-$0.4) and the remaining 12 as consistent with steep intrinsic power-law spectra typical of RSNe or SNRs. We find no difference in the source classification in our study.

The flux density for the sources with spectra consistent with RSNe or SNRs, ranges from 2 - 30 mJy at 5~GHz. The relationship between the rise time and peak luminosity for core-collapse supernovae by \citet{CFN06} predicts a rise time $\leq$110 days. A typical RSNe follows a power law decay that begins after this fast rise, and would be detectable over many years. However, no such decay has been observed for these compact sources over two decades, suggesting that they are probably very old SNRs.

%However, the spectra for compact sources in NGC~253 are largely inconsistent with H~\textsc{ii} regions (as are the lower limits in brightness temperature, $\sim10^{5}$ K), but are instead consistent with synchrotron sources (such as RSNe or SNRs) affected by foreground free-free absorption. 
%Thus, 

\placefigure{fig:ffspectra}
\placetable{tab:ffparams}

%----------------------------------------------------------------------------------------------------------------------------------------------------------------------------------
\subsection{The Supernova Remnant 5.48-43.3}
\label{sec:548}

\subsubsection{Morphology and Small-scale Features}
\label{sec:548M}
%-------------------------------------------------

\begin{figure*}[htbp]\scriptsize
\centering
\subfloat{\includegraphics[angle=-90,scale=0.3,origin=c]{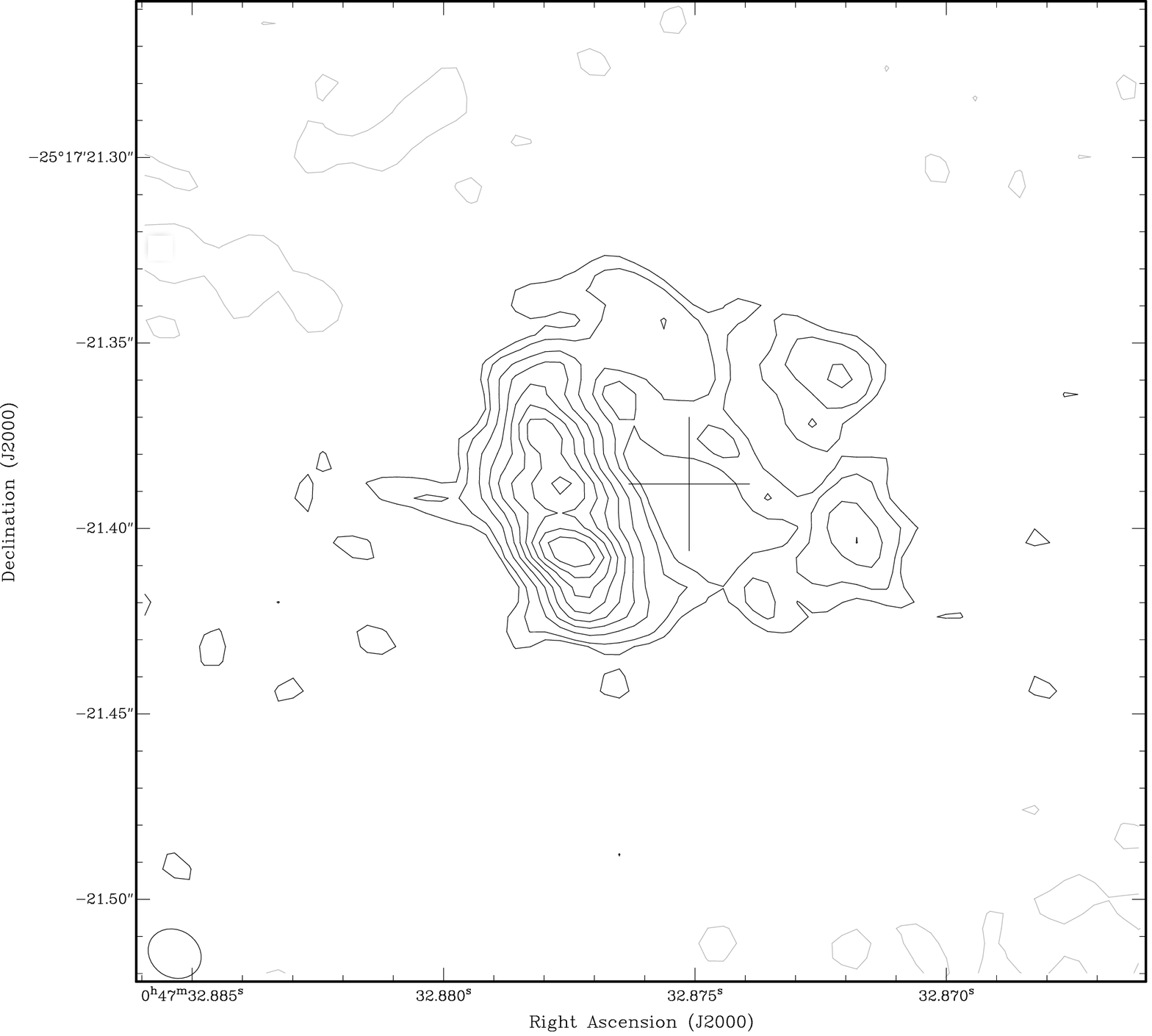}}
\subfloat{\includegraphics[angle=-90,scale=0.3,origin=c]{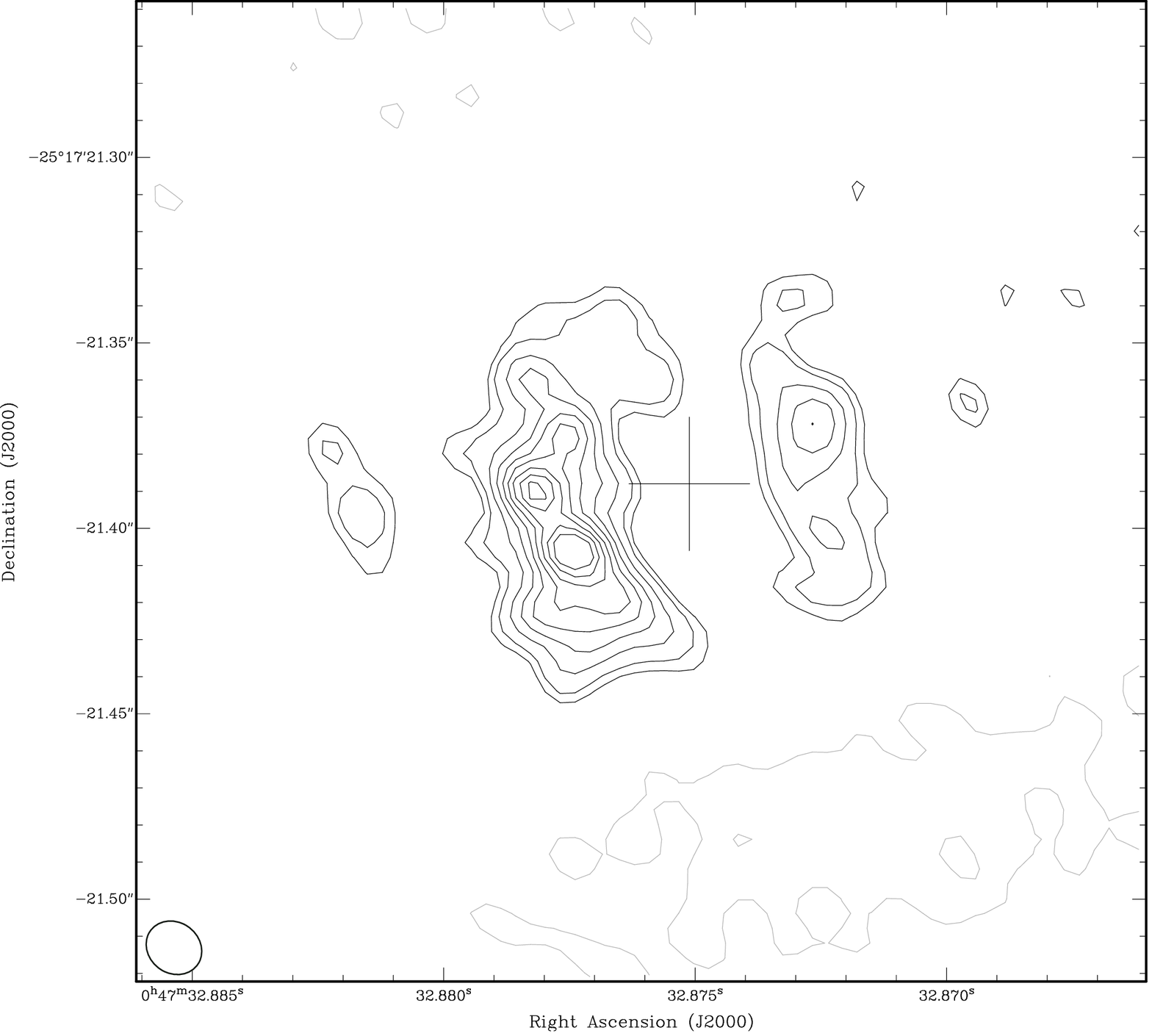}}
\caption{LBA images of 5.48-43.3 at 2.3~GHz taken on  from 2004 April 16/17 (left) from LT06 and 2008 June 5/6 (right). The contours in both images corresponds to $-$15$\%$, 15, 20$\%$, 30$\%$, 40$\%$, 50$\%$, 60$\%$, 70$\%$, 80$\%$, and 90$\%$ of the 2004 peak (2.4~\mjybm). Both images are restored with the same beam, 13$\times$15 mas with a position angle of 58\deg. The cross indicates the assumed source centre used to obtain the radial flux profile in \Fig~\ref{fig:iring}.}
\label{fig:548v2}
\end{figure*}
%----------------------------------------------
The brightest source in NGC~253 at 2.3~GHz, 5.48-43.3, is the only fully resolved SNR in the starburst galaxy with the full LBA array (i.e. including the long baselines of Cd and Ho). In \fig~\ref{fig:548v2} we present the two high sensitivity VLBI images of 5.48-43.3 (the 2004 and 2008 epochs), restored with a beam size of  13$\times$15 mas and beam position angle of 58\deg. The contours were chosen to represent identical surface brightness levels in both images.

As first noted by LT06, 5.48$-$43.3 appears to be a shell-type SNR, with a diameter of $\sim$70 - 90 mas (1.4 - 1.8 pc). At both epochs, the structure of 5.48$-$43.3 is dominated by the eastern lobe, which has a higher flux density than the western lobe by a factor of 3. This may be the result of interactions with a denser interstellar medium in the direction of the eastern lobe.

Comparing the epochs, we notice that there are  several differences in the lobes. Such apparent changes in the small scale structures are possibly due to ambiguities caused by the combination of structural evolution and image fidelity limitations, resulting from incomplete sampling of the $uv$-plane. The effects of $uv$-plane sampling on the appearance and evolution of complex small-scale structures, within spherically symmetric shell-like sources, are well documented by \citet{Heywoodetal2009}; they demonstrated that sparse $uv$-plane sampling, and the non-uniqueness of deconvolution, can add complex azimuthal structure to a radio brightness distribution that is in reality morphologically simple. 

One striking feature of the \citet{Heywoodetal2009} results is the departure from a spherically symmetric shell to a two-lobed structure (similar to SN~1987A \citep{NGetal2011} and 5.48$-$43.3) when the $uv$-plane coverage lacks intermediate baselines, similar to the $uv$ coverage of both the 2004 (see \fig~2 of LT06) and the 2008 (\fig~\ref{fig:uvcov}) observations. Thus, it is possible that 5.48-43.3 possesses a spherically symmetric (or slightly elliptical) morphology that is not recovered due to low sensitivity associated with sparse sampling of the $uv$-plane. This effect is noticeable in the epochs without the intermediate baselines to Tidbinbilla (2006 and 2007), with the non-detection of the weaker, western lobe.

\subsubsection{Expansion of 5.48-43.3}

\begin{figure}[htp]
\centering
\includegraphics[scale=0.4]{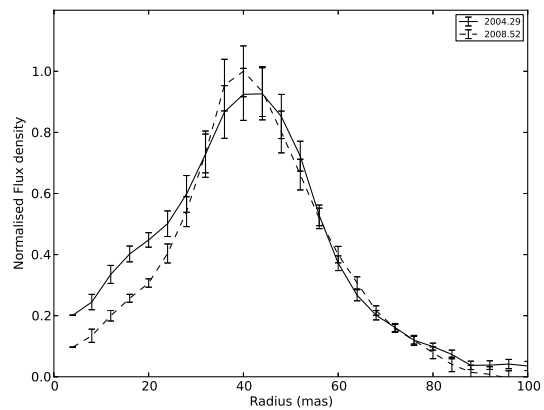} 
\caption{ Radial flux density profile of the SNR 5.48-43.3 for the 2004 (solid line) and 2008 (dashed line) observations. }
\label{fig:iring}
\end{figure}

With multiple high resolution observations at the same frequency, it may be possible to determine the expansion speed of 5.48-43.3. In order to carry out such measurements, we use the \textsc{aips} task \textsc{iring} to measure radial profiles (averaged in azimuth) of the source at each epoch. This method has been successfully implemented to determine the expansion speed of resolved SNRs in M82 \citep{Beswicketal06, Fenechetal10}. Owing to differences in the small-scale structure between the two epochs (as discussed in the previous section), determination of the geometrical centre of 5.48-43.3 is difficult. To account for any positional offset between the epochs, the position of the peak surface brightness in the 2008 image was aligned to the position of the 2004 peak surface brightness, using the \textsc{aips} task \textsc{ogeom}. A common geometrical centre for both images was estimated by visual inspection, and indicated by the crosses in \Fig~\ref{fig:548v2}. Using this position as the reference point, the radial profiles were obtained by measuring the integrated flux density within a series of 4~mas thick annuli, and are plotted in \fig~\ref{fig:iring}. No discernible expansion between the two epochs can be seen in \Fig~\ref{fig:iring}. At the smaller radii, higher integrated flux densities were recovered for the 2004 image. This is possibly due to differences in the short spacings of the $uv$ coverages and/or deeper cleaning in the 2008 image.

%------------------------------------------
%----------------------------------

If we consider an upper limit of the expansion speed on $\nu\,=\,10^{4}$~km~s$^{-1}$ \citep{Beswicketal06,Fenechetal10} for SNRs in NGC~253, an expansion of $\sim$2.2~mas, $\sim15\%$ of the 2.3~GHz LBA beam, is expected for 5.48$-$43.3 over 4.3 years between the 2004 and 2008 LBA observations. To measure the expansion of 5.48$-$43.3, multiple observations separated by $>$10 years with the current LBA baselines, or observations with longer baselines ($>$3000~km) separated by $>$5 years would be needed. 
 %----------------------------------------------------------------------------------------------------------------------------------------------------------------------------------

\subsection{The Supernova Rate in NGC~253}
\label{sec:SR}
  
\subsubsection{Lower limit estimation} 
 
 A lower limit on the supernova rate can be estimated based on the number of detected SNRs, their size, and an assumed expansion rate (LT06). Based upon the method by \citet{UA94} and an assumed radial expansion rate of $10^{4}$~km~s$^{-1}$,  LT06 estimated the lower limit of the  supernova rate to be $0.14(v/10^{4}$~km~s$^{-1})$. Our results from \Sect~\ref{sec:548} are consistent with this as the minimum supernova rate.
 
\subsubsection{Previous estimates of the supernova rate upper limit}
 
No new sources have been detected in NGC~253 after two decades of multi-wavelength, high resolution radio observations. The consequences of the non-detections for estimates of the supernova rate were first investigated by \citet[hereafter UA91]{UA91}. They assumed a hypothetical population of RSNe whose flux densities at 5~GHz peak 100 days after the optical maxima and decay as the $-$0.7 power of time \citep{Weileretal1986,Weileretal89}. The RSNe population was assumed to have peak luminosities uniformly distributed between 5 and 20 times that of Cas~A, and an NGC~253 distance of 2.5~Mpc (TH85) was also assumed. UA91 found that $\sim$2/3 of the RSNe that occurred during the 18 month period between two 5~GHz epochs should be detectable above the second epoch's sensitivity limit. By assuming the events are Poisson distributed in time, UA91 determined with 95$\%$ confidence an upper limit to the supernova rate of 3.0~yr$^{-1}$ in NGC253. This model was used to estimate the supernova rate upper limit for two subsequent 5~GHz epochs, $\sim$1.4~yr$^{-1}$ for a third epoch 2.5 years later \citep{UA94} and  $\sim$0.3~yr$^{-1}$ for a fourth epoch, 4.0 years later (UA97). 

With new distance measurements for NGC~253 (3.94$\pm$0.5~Mpc, \citealt{Karachentsevetal2003}) and two additional high resolution observations, LT06 developed a new model based on the principles of UA91 and using new data: a 5~GHz\footnote{LT06 erroneously lists a 5.0~GHz observation in place of the 8.3~GHz observation by \citet{Mohanetal2005}.} VLA observation by \citet{Mohanetal2005}; and a 2.3~GHz LBA observation by LT06. The RSNe at 2.3~GHz were assumed to peak 200 days after the optical maxima at the same flux density as at 5~GHz (see \citealt{Weileretal1986}). LT06 also considered the effects of free-free absorption by adjusting the flux densities of the RSNe with \eqn~\eqref{eq:spectra2}, assuming a median value of $\tau_{0}$ = 6. The number of RSNe that occur between epochs was determined via a Poisson-distributed random number, given a specified supernova rate. The radio luminosities of the RSNe were allowed to evolve using the parametric equation of \citep{Weileretal1986} into supernovae remnants (SNRs) and then, given the time and sensitivity of the observation, a test was made to determine if each SNR could be detected. A Monte Carlo simulation was used to drive the model ($\sim$ 10000 iterations), where the proportion of SNRs detected, $\beta_{SN}$ at the end of each epoch was obtained. The simulation was seeded with an initial supernova rate of 0.1~yr$^{-1}$ producing a confidence level for the $\beta_{SN}$ at each epoch. Using linear interpolation, the supernova rate required to drive the simulation to a 95$\%$ confidence limit was determined and the simulation was repeated until the confidence limit was achieved. The resulting supernova rate was then used to seed the next epoch and the process was continued until all epochs were processed.

Using this model, LT06 found that the increased distance, combined with the effects of free-free absorption, decreased $\beta_{SN}$ which resulted in an upper limit of the supernova rate at the end of the final epoch of 2.4~yr$^{-1}$. A summary of the model parameters used by \citet{UA91, UA94, UA97} and LT06 to derive upper limits on the supernova rate are listed in \Tab~\ref{tab:SNRmodelparams}.

%---------------------------------------------

\begin{table}[htbp]\scriptsize
\begin{minipage}{8cm}
\renewcommand\footnoterule{}
\caption{\textsc{Summary of input parameters to different supernova rate models of ngc~253}}
\label{tab:SNRmodelparams}
\medskip
\hfill{}
{\renewcommand{\tabcolsep}{3.5pt}
\begin{tabular}{lccc}
\hline
\hline
Input Parameters & UA$^{a}$ & LT06 & This Paper \\ 
\hline
 No. of Epochs \dotfill &  4 & 6 & 11  \\ 
 $\Delta$F$^{b}$ \dotfill & $\mathcal{U}$(5,20)$^{c}$ & $\mathcal{U}$(5,20) & Log-Gaussian  \\ 
Frequency, $\nu$ (GHz) \dotfill & 5 & 2.3 $\&$ 5 & 2.3, 5, 8.3, 15 $\&$ 23  \\ 
 $T_{r}$ $^{d}$ (days) \dotfill & 100 & 200 $\&$ 100 & 200, 100, 100, 70 $\&$ 50   \\
 $\beta^{e}$ \dotfill & $-$0.7 & $-$0.7 & $-$0.7 \\
 Spectral Index, $\alpha$ \dotfill & $\cdots$ & 0 & time-dependent$^{f}$\\
 $\tau_{0}$ $^{g}$ \dotfill & $\cdots$ & 6$^{h}$ & 6 \\
 Distance (Mpc) \dotfill & 2.55 & 3.94 & 3.44 \\
\hline
\\
\multicolumn{4}{l}{$^{a}$\citet{UA91, UA94, UA97}}\\
\multicolumn{4}{l}{$^{b}$Type II Supernovae Peak Luminosity Distribution in}\\
 \multicolumn{4}{l}{units of Cas~ A luminosity at 5~GHz =  $7.34~\times~10^{24}$~ergs~s$^{-1}$~Hz$^{-1}$}\\
 \multicolumn{4}{l}{$^{c}\mathcal{U}$(a,b) means a uniform distribution with limits, a $\&$ b}\\
 \multicolumn{4}{l}{$^{d}$Time after optical peak to reach radio peak luminosity at $\nu$~GHz}\\
 \multicolumn{4}{l}{ listed above.}\\
 \multicolumn{4}{l}{$^{e}$RSNe luminosity decay with time \citep{Weileretal1986,Weileretal89}}\\
 \multicolumn{4}{l}{$^{f}\alpha^{5}_{2.3}$ from \citet{Weileretal1986}, $\alpha^{23}_{5}$ from \citet{Weileretal2007}}\\
 \multicolumn{4}{l}{$^{g}$Free-Free Optical Depth at 1~GHz}\\
 \multicolumn{4}{l}{$^{h}$Median value based upon the results of LT06}\\
\end{tabular}
}
\hfill{}
\vspace{-2ex} 
\end{minipage}
\end{table}

%-----------------------------------------------------------------------------------------

\subsubsection{Determining a new supernova rate upper limit}

\begin{figure}[htp]
\centering
\includegraphics[trim=1cm 0cm 0cm 0cm,clip=true,scale=0.45]{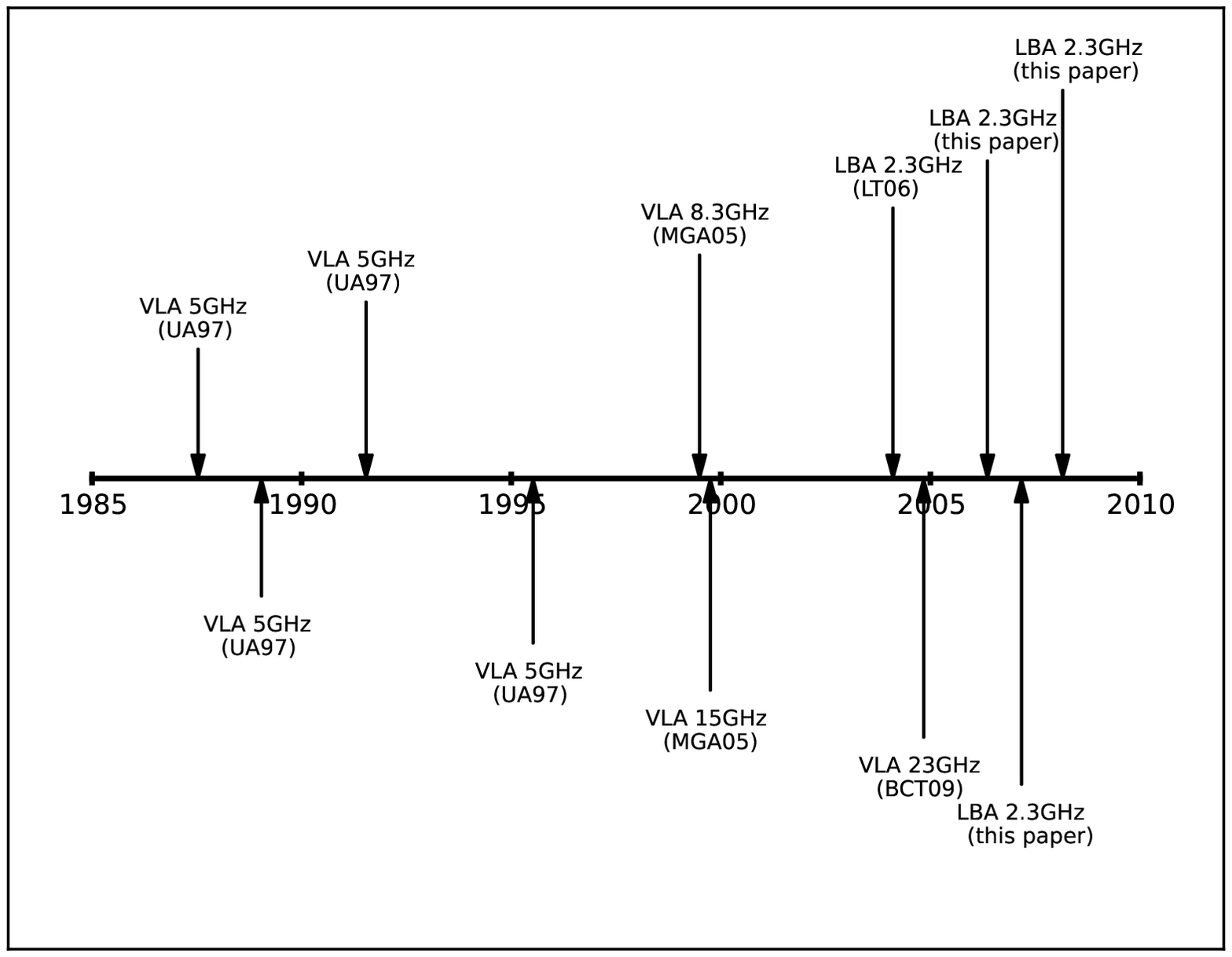} 
\caption{Time-line of observations used to estimate the upper limits on the supernova rate in NGC~253 by  \citet{UA91, UA94, UA97}, LT06 and this paper.  The new observations introduced in the model by this paper are the VLA~8.3~GHz $\&$ 15~GHz (MGA05), VLA 23~GHz (BCT09) and three LBA 2.3~GHz (this paper) observations. \textsc{note}: MGA05 - \citet{Mohanetal2005}; BCT09 - \citet{Brunthaleretal2009} }
\label{fig:timeline}
\end{figure}

With additional high resolution radio observations of NGC~253 since LT06, improved distance measurements to NGC~253, and availability of a more realistic RSNe luminosity function \citep{Lienetal2011}, it is worth revisiting the estimates of the upper limit on the supernova rate. The additional NGC~253 observations are the three new 2.3~GHz LBA observations in this paper, plus 23~GHz VLA observations by \citet{Brunthaleretal2009} and 15~GHz VLA observation by \citet{Mohanetal2005}. \Fig~\ref{fig:timeline} gives a time-line of all observations of NGC~253 used to determine the supernova rate. To further constrain the upper limit on the supernova rate with these observations, we have developed a new model based on the principles of \citet{UA91} and LT06, as well as incorporating suggested improvements by LT06. 

%------------------------------
%------------------------------

%----------------------------------------------------
\begin{center}
%\begin{savenotes}
\begin{table*}[htp]\scriptsize
%\fbox{% Remove this line to delete the frame
\begin{minipage}{18cm}
\renewcommand\footnoterule{}
%{\small
\caption{\textsc{observed 5~GHz peak flux density of core-collapse supernova}}
\label{tab:SNRflux}
\medskip
\hfill{}
%{\renewcommand{\tabcolsep}{2pt}
\begin{tabular}{llccccccc}
\hline
\hline
\noalign{\smallskip}
Name & Galaxy & Type & Distance$^{a}$ &  Detected?$^{b}$ &  S$_{5}$ $^{c}$&  L$_{6}$ $^{d}$ & L$_{5;CasA}$  $^{d}$   & Ref. \\
 &  &  & (Mpc) & & (mJy) & &  & \\
\hline
 %\endhead
SN1983N & M83 & Ibc & 4.5 & D & 40.10 & 9.72 & 132.45 & 1 \\ 
SN1984L & NGC 4991 & Ibc & 21.7 & D & 4.59 & 25.86 & 352.54 & 2 \\ 
SN1990B & NGC 4568 & Ibc & 18.2 & D & 1.26 & 4.99 & 68.08 & 3 \\ 
SN1994I & M51 & Ibc & 8.0 & D & 16.20 & 12.50 & 170.38 & 4 \\ 
SN1998bw & ESO184-G082 & Ibc & 37.3 & D & 37.40 & 622.59 & 8487.24 & 5 \\ 
SN2000C & NGC 2415 & Ibc & 59.0 & E & 0.19 & 7.79 & 106.18 & 6 \\ 
SN2001B & IC391 & Ibc & 24.0 & D & 2.56 & 17.64 & 240.51 & 6 \\ 
SN2001ci & NGC 3079 & Ibc & 17.0 & D & 1.06 & 3.67 & 50.06 & 6 \\ 
SN2002ap & NGC 628 & Ibc & 7.3 & D & 0.45 & 0.29 & 3.94 & 7 \\ 
SN2002cj & ESO582-G5 & Ibc & 106.0 & D & 0.22 & 29.58 & 403.19 & 6 \\ 
SN2002dg & Anon. & Ibc & 215.0 & E & 0.09 & 50.88 & 693.65 & 6 \\ 
SN2003L & NGC 3506 & Ibc & 92.0 & D & 2.56 & 258.75 & 3527.31 & 8 \\ 
SN2003bg & MCG-05-10-15 & Ibc & 19.6 & D & 52.81 & 242.74 & 3309.07 & 9 \\ 
SN2004cc & NGC 4568 & Ibc & 18.0 & E & 3.26 & 12.64 & 172.28 & 10 \\ 
SN2004dk & NGC 6118 & Ibc & 23.0 & D & 1.80 & 11.39 & 155.31 & 10 \\ 
SN2004gq & NGC 1832 & Ibc & 26.0 & D & 5.75 & 46.51 & 634.00 & 10 \\ 
SN2007bg & Anon. & Ibc & 152.0 & D & 1.10 & 304.91 & 4156.62 & 11 \\ 
SN2007gr & NGC 1058 & Ibc & 7.7 & D & 1.00 & 0.71 & 9.67 & 12 \\ 
SN2008D & NGC 2770 & Ibc & 27.0 & E & 3.00 & 26.17 & 356.72 & 13 \\ 
SN2009bb & NGC 3278 & Ibc & 40.0 & D & 12.00 & 229.73 & 3131.69 & 14 \\ 
SN2010as & NGC 6000 & Ibc & 27.3 & D & 1.40 & 12.48 & 170.19 & 15 \\ 
PTF11qcj & SDSS$^{f}$ & Ib & 124.0 & D & 5.43 & 998.98 & 13618.22 & 16 \\ 
\hline \hline
SN1970g & NGC 5457 & II & 7.2 & D & 2.50 & 1.55 & 21.14 & 17 \\ 
SN1974E & NGC 4038 & II? & 20.9 & D & 1.24 & 6.48 & 88.35 & 18 \\ 
SN1978K & NGC 1313 & II & 4.1 & D & 518.00 & 104.19 & 1420.28 & 18 \\ 
SN1979C & NGC 4321 & IIL & 22.0 & D & 12.40 & 71.81 & 978.91 & 18 \\ 
SN1980K & NGC 6946 & IIL & 7.0 & D & 2.45 & 1.44 & 19.58 & 18 \\ 
SN1981k & NGC 4258 & II & 6.6 & D & 5.15 & 2.68 & 36.59 & 18 \\ 
SN1982aa & NGC 6052 & II & 66.0 & D & 19.10 & 995.48 & 13570.58 & 18,19 \\ 
SN1985L & NGC 5033 & II & 12.9 & D & 1.56 & 3.11 & 42.34 & 20 \\ 
SN1986E & NGC 4302 & II & 16.8 & D & 0.33 & 1.11 & 15.19 & 21 \\ 
SN1986J & NGC 891 & IIn & 12.0 & D & 135.00 & 232.60 & 3170.83 & 18 \\ 
SN1988Z & MCG+03-28-022 & IIp & 94.5 & D & 1.85 & 197.67 & 2694.71 & 22 \\ 
SN1993J & M81 & IIb & 3.6 & D & 96.90 & 15.28 & 208.26 & 23 \\ 
SN1995N & MCG-02-38-17 & IIn & 24.0 & E & 9.00 & 62.03 & 845.56 & 24 \\ 
SN1997eg & NGC 5012 & II & 40.0 & E & 10.00 & 191.44 & 2609.74 & 25 \\ 
SN1998S & NGC 3877 & II & 17.0 & E & 1.04 & 3.60 & 49.02 & 26 \\ 
SN1999em & NGC 1637 & II & 7.8 & E & 0.30 & 0.22 & 2.98 & 26 \\ 
SN2000ft & NGC 7469 & II? & 70.0 & D & 1.76 & 103.30 & 1408.25 & 27 \\ 
SN2001ig & NGC 7424 & IIb & 11.5 & D & 21.90 & 34.65 & 472.41 & 28 \\ 
SN2001gd & NGC 5033 & IIb & 20.0 & D & 7.96 & 38.10 & 519.34 & 29 \\ 
SN2004dj & NGC 2403 & IIp & 3.6 & E & 1.90 & 0.29 & 4.02 & 30 \\ 
SN2004et & NGC 6946 & II & 5.5 & E & 2.50 & 0.90 & 12.34 & 31 \\ 
SN2006jd & UGC 4179 & IIb & 76.8 & D & 2.22 & 156.39 & 2131.91 & 32 \\ 
SN2008ax & NGC 4490 & IIp & 8.1 & D & 4.00 & 3.13 & 42.70 & 33 \\ 
SN2008iz & M82 & II & 3.2 & E & 180.00 & 22.05 & 300.64 & 34 \\ 
SN2011cb & PGC 69707 & IIb & 29.2 & D & 0.73 & 7.46 & 101.73 & 35 \\ 
SN2011dh & M51 & IIp & 8.0 & D & 7.80 & 6.02 & 82.04 & 36 \\ 
SN2011ei & NGC 6925 & IIb & 28.5 & D & 0.56 & 5.48 & 74.72 & 37 \\ 
A25 & Arp220 & IIp & 44.8 & E & 0.30 & 7.20 & 98.21 & 38 \\ 
A27 & Arp220 & IIp? & 44.8 & E & 0.75 & 18.01 & 245.52 & 38 \\ 
\hline
 &  &  &  &  &  &  &  & \\ 

\multicolumn{9}{l}{ Ref: 1.  \citet{Srameketal1984}; 2. \citet{Weileretal1986}; 3. \citet{vanDyketal1993}; 4. \citet{Weileretal2011};
}\\
\multicolumn{9}{l}{ 5. \citet{Kulkarnietal1998}; 6. \citet{Bergeretal2003}; 7. \citet{Bergeretal2002}; 8. \citet{Soderbergetal2005}; 
}\\
\multicolumn{9}{l}{9. \citet{Soderbergetal2006}; 10. \citet{Wellonsetal2012}; 11. \citet{Salasetal2013}; 12. \citet{Soderbergetal2010}; 
}\\
\multicolumn{9}{l}{13. \citet{Soderbergetal2008}; 14. \citet{Bietenholzetal2010}; 15. \citet{Ryderetal2010};  16. \citet{Corsietal2013}; 
}\\
\multicolumn{9}{l}{ 17. \citet{Weileretal89}; 18. \citet{Weileretal2002}; 19.  \citet{Metcalfeetal2005}; 20. \citet{vanDyketal1998}; 
}\\
\multicolumn{9}{l}{ 21.  \citet{Montesetal1997}; 22. \citet{Williamsetal02}; 23. \citet{Weileretal2007}; 24.  \citet{Chandraetal09}; 
}\\
\multicolumn{9}{l}{ 25. \citet{Laceyetal1998}; 26. \citet{Pooleyetal2002}; 27. \citet{PT2009}; 28. \citet{Ryderetal2004}; 
}\\
\multicolumn{9}{l}{ 29. \citet{Stockdaleetal07};  30. \citet{Beswicketal05}; 31. \citet{Argoetal2005}; 32. \citet{Chandraetal12}; 
}\\
\multicolumn{9}{l}{33. \citet{Romingetal09};  34. \citet{Gendreetal2012}; 35. \citet{Kraussetal2012}; 36.  \citet{Ryderetal2011a}; 
}\\
\multicolumn{9}{l}{ 37.  \citet{MMS2013}; 38. \citet{Bondietal2012}.
}\\
\noalign{\smallskip}
\cline{1-4}\\
\multicolumn{9}{l}{$^{a}$Estimated from NASA/IPAC Extragalactic Database (NED), where appropriate}\\
\multicolumn{9}{l}{$^{b}$D - peak flux detected at 5~GHz; E - peak flux at 5~GHz estimated} \\
\multicolumn{9}{l}{$^{c}$Peak Flux Density at 5~GHz} \\
\multicolumn{9}{l}{$^{d}$Peak luminosity at 5~GHz; units = $10^{26}$~ergs~s$^{-1}$~Hz$^{-1}$}\\
\multicolumn{9}{l}{$^{e}$Luminosity ratio to Cassiopeia~A (Cas~A) 5~GHz peak luminosity: $7.34~\times~10^{24}$~ergs~s$^{-1}$~Hz$^{-1}$}\\
\multicolumn{9}{l}{$^{f}$SDSS J131341.57+471757.2}\\
\end{tabular}
%}
\hfill{}
 \vspace{-2ex}
 \end{minipage}%
 % }
\end{table*}
%\end{savenotes}
\end{center}
%--------------------------------------------------------------------------------------

%----------------------------------------------------
\begin{center}
%\begin{savenotes}
\begin{table*}[htp]\scriptsize
%\fbox{% Remove this line to delete the frame
\begin{minipage}{18cm}
\renewcommand\footnoterule{}
%{\small
 \caption{\textsc{peak flux density upper limits of undetected core-collapse supernova at 5~GHz}}
\label{tab:SNRupperlims}
\medskip
\hfill{}
\begin{tabular}{llccccccc}
\hline
\hline
\noalign{\smallskip}
Name & Galaxy &  Type  &  Distance & S$_{5}$& L$_{5}$& L$_{5;CasA}$ &  p-value$^{a}$ &Ref.\\

&  &  & (Mpc) &  (mJy) &  &  &  &  \\ 
\hline
SN1980I & NGC 4374 & Ibc & 22.0 & $<$0.50 & $<$2.90 & $<$39.47 & 0.27 & 1,2 \\ 
SN1980N & NGC 1316 & Ibc & 33.0 & $<$0.60 & $<$7.82 & $<$106.58 & 0.39 & 1 \\ 
SN1981B & NGC 4536 & Ibc & 33.0 & $<$0.30 & $<$3.91 & $<$53.29 & 0.30 & 1 \\ 
SN1999ex & IC 5179 & Ibc & 54.0 & $<$0.07 & $<$2.48 & $<$33.77 & 0.25 & 3 \\ 
SN2000cr & NGC 5395 & Ibc & 54.0 & $<$0.04 & $<$1.47 & $<$19.98 & 0.20 & 3 \\ 
SN2000ew & NGC 3810 & Ibc & 15.0 & $<$0.07 & $<$0.18 & $<$2.46 & 0.06 & 3 \\ 
SN2000fn & NGC 2526 & Ibc & 72.0 & $<$0.05 & $<$2.79 & $<$38.05 & 0.26 & 3 \\ 
SN2001M & NGC 3240 & Ibc & 56.0 & $<$0.03 & $<$1.05 & $<$14.32 & 0.17 & 3 \\ 
SN2001ai & NGC 5278 & Ibc & 118.0 & $<$0.05 & $<$8.16 & $<$111.29 & 0.39 & 3 \\ 
SN2001bb & IC 4319 & Ibc & 72.0 & $<$0.06 & $<$3.72 & $<$50.73 & 0.29 & 3 \\ 
SN2001ef & IC 381 & Ibc & 38.0 & $<$0.12 & $<$1.99 & $<$27.09 & 0.23 & 3 \\ 
SN2001ej & UGC 3829 & Ibc & 63.0 & $<$0.04 & $<$2.09 & $<$28.48 & 0.23 & 3 \\ 
SN2001is & NGC 1961 & Ibc & 60.0 & $<$0.07 & $<$3.02 & $<$41.10 & 0.27 & 3 \\ 
SN2002J & NGC 3464 & Ibc & 58.0 & $<$0.09 & $<$3.42 & $<$46.64 & 0.28 & 3 \\ 
SN2002bl & UGC 5499 & Ibc & 74.0 & $<$0.05 & $<$2.95 & $<$40.19 & 0.27 & 3 \\ 
SN2002bm & MCG013219 & Ibc & 85.0 & $<$0.07 & $<$5.71 & $<$77.78 & 0.35 & 3 \\ 
SN2002cg & UGC 10415 & Ibc & 150.0 & $<$0.07 & $<$19.92 & $<$271.58 & 0.51 & 3 \\ 
SN2002cp & NGC 3074 & Ibc & 80.0 & $<$0.04 & $<$2.68 & $<$36.54 & 0.26 & 3 \\ 
SN2002dn & IC 5145 & Ibc & 115.0 & $<$0.04 & $<$6.80 & $<$92.76 & 0.37 & 3 \\ 
SN2002ge & NGC 7400 & Ibc & 47.0 & $<$0.13 & $<$3.44 & $<$46.84 & 0.29 & 3 \\ 
SN2002gy & UGC 2701 & Ibc & 114.0 & $<$0.04 & $<$6.06 & $<$82.67 & 0.35 & 3 \\ 
SN2002hf & MCG05320 & Ibc & 88.0 & $<$0.04 & $<$3.71 & $<$50.52 & 0.29 & 3 \\ 
SN2002hn & NGC 2532 & Ibc & 82.0 & $<$0.04 & $<$3.54 & $<$48.26 & 0.29 & 3 \\ 
SN2002ho & NGC 4210 & Ibc & 42.0 & $<$0.05 & $<$1.14 & $<$15.54 & 0.17 & 3 \\ 
SN2002hy & NGC 3464 & Ibc & 58.0 & $<$0.07 & $<$2.98 & $<$40.60 & 0.27 & 3 \\ 
SN2002hz & UGC 12044 & Ibc & 85.0 & $<$0.03 &$<$2.51 &$<$34.18 & 0.25 & 3 \\ 
SN2002ji & NGC 3655 & Ibc & 23.0 & $<$0.04 & $<$0.27 & $<$3.71 & 0.08 & 3 \\ 
SN2002jj & IC 340 & Ibc & 66.0 & $<$0.03 & $<$1.72 & $<$23.45 & 0.21 & 2 \\ 
SN2002jp & NGC 3313 & Ibc & 58.0 & $<$0.04 & $<$1.53 & $<$20.85 & 0.20 & 3 \\ 
SN2002jz & UGC 2984 & Ibc & 24.0 & $<$0.03 & $<$0.17 & $<$2.35 & 0.06 & 3 \\ 
SN2003jd & MCG0159021 & Ibc & 81.0 & $<$0.05 & $<$4.08 & $<$55.65 & 0.31 & 4 \\ 
SN2011hp & NGC 4219 & Ibc & 23.7 & $<$0.20 & $<$1.34 & $<$18.32 & 0.19 & 5 \\ 
SN2005cs & M51 & II & 8.0 & $<$0.37 & $<$0.29 & $<$3.89 & 0.08 &6 \\ 
SN2009mk & PGC 474 & IIb & 18.3 & $<$0.15 & $<$0.60 & $<$8.19 & 0.12 & 7 \\ 

\hline
 & & &  &  &  &  & &  \\ 
\multicolumn{9}{l}{ Ref: 1. \citet{Weileretal89}; 2. \citet{Weileretal1986}; 3 \citet{Bergeretal2003};4. \citet{Soderbergetal2006a};  }  \\
\multicolumn{9}{l}{ 5. \citet{Ryderetal2011}; 6. \citet{Stockdaleetal2005}; 7. \citet{Ryderetal2010b} }\\
\noalign{\smallskip}
\cline{1-4}\\
\multicolumn{9}{l}{Notes: - Where appropriate symbols and headings have same meaning as \Tab~6. Only core-collapse}\\
\multicolumn{9}{l}{ supernovae that were observed in the radio within 200 days of the optical maximum are selected.} \\
\noalign{\smallskip}
\multicolumn{9}{l}{$^{a}$The probability the supernova's peak radio luminosity is actually less than the observed}\\
 \multicolumn{9}{l}{ upper limits given the cumulative distribution function derived from Figure 12.}
\end{tabular}
\hfill{}
\vspace{-2ex}
 \end{minipage}%
 % }
\end{table*}
%\end{savenotes}
\end{center}
%--------------------------------------------------------------------------------------

\subsubsection*{Improved core-collapse supernovae peak luminosity distribution}

At 5~GHz and 2.3~GHz our hypothetical RSN is the same as that used by UA91 and LT06 (see previous section and \Tab~\ref{tab:SNRmodelparams}). At 8.3~GHz, 15~GHz and 23~GHz we assume that the RSN luminosity peaks 100, 70 and 50 days after the optical maximum at the same luminosity as at 5~GHz. This assumption closely agrees with the light curves of SN~1993J  \citep{Weileretal2007}.

In the models of UA91 and LT06, the distribution of the 5~GHz peak luminosities of RSNe was assumed to be uniformly distributed between 5 and 20 times the luminosity of Cas~A. This assumption follows \citet{Weileretal89}, who came to this conclusion after comparing the peak flux densities at 5~GHz of 16 Type II supernovae (three detections and 13 upper limits). Since the study by \citet{Weileretal89} there have been many more new detections of Type II as well as Type Ibc supernovae in the radio. Thus it is worth re-investigating the  5~GHz peak luminosity distribution of these objects.

%------------------------------------
\begin{figure}[h]
\centering
\includegraphics[trim=1cm 0cm 0cm 0cm,clip=true,scale=0.42]{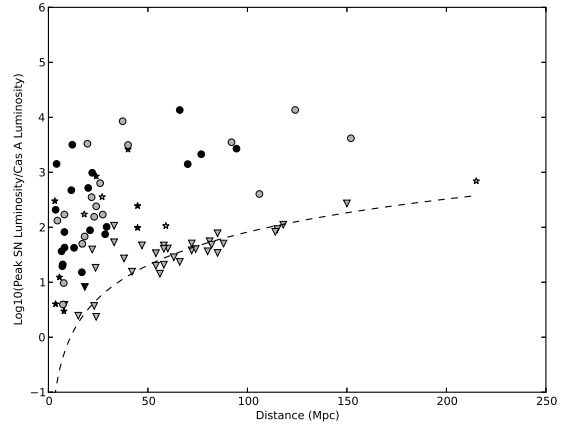}
\caption{The 5~GHz peak luminosity ratio to Cas~A as a function of distance for Type II supernovae (black points) and Type Ibc supernovae (grey points). The circles and stars are supernovae with detected and estimated peak luminosities (29 Type II and 22 Type Ibc, see \Tab~\ref{tab:SNRflux}). The upside down triangles are supernovae whose peak luminosities have not been detected (2 Type II and 32 Type Ibc, see \Tab~\ref{tab:SNRupperlims}). The dashed lines is the average 3$\sigma$ detection limit of the VLA from \citealt{Bergeretal2003}. }
\label{fig:snedist}
\end{figure}
%------------------------------------

\begin{figure}[h]
\centering
\includegraphics[trim=1cm 0cm 0cm 0cm,clip=true,scale=0.42]{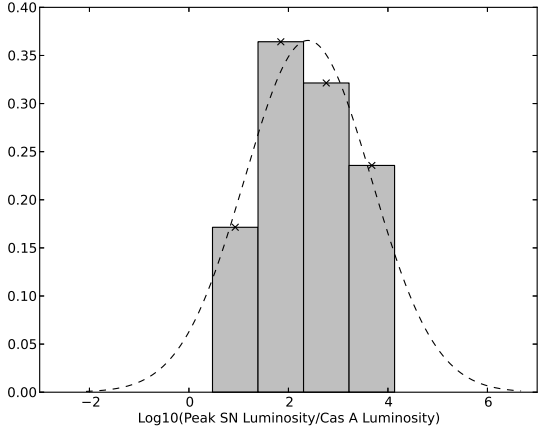}
\caption{Normalised radio luminosity distribution at 5~GHz of 51 core-collapse supernovae as a function of log$_{10}$(L$_{5;CasA}$). The x-axis mirrors the y-axis of \Fig~\ref{fig:snedist} and is given in \Tab~\ref{tab:SNRflux}. The data are binned to $\Delta$log$_{10}$(L$_{5;CasA}$) = 1. The dashed curve shows a non-linear least-squares fitted Gaussian to the underlying data (black x's). The resulting fit: mean log$_{10}$(L$_{5;CasA}$)= 2.39, $\sigma$ = 1.27 and $\chi^{2}$ = 0.018.  }
\label{fig:snehist}
\end{figure}

Table~\ref{tab:SNRflux} lists 51 core-collapse (29 Type~II $\&$ 22 Type~Ibc) supernovae whose 5~GHz peak flux density have either been detected or estimated via spectral index\footnote{For peak flux densities detected at 8.4~GHz, a spectral index of 0 is assumed, following SN1993J \citep{Weileretal2007}} or extrapolation using the models by \citet{Chevalier1982} or \citet{Weileretal2002}. This table is represented graphically in \Fig~\ref{fig:snedist}, where the Type Ibc (grey symbols) and Type II (black symbols) luminosity ratio to Cas~A's at 5~GHz is plotted on the y-axis and their distances on the x-axis. Included in the plots are the 5~GHz peak luminosity upper limits  for 34 core-collapse supernovae not detected in the radio (\Tab~\ref{tab:SNRupperlims}), that follow the average 3$\sigma$ detection limit of the VLA (the dashed lines, from \citealt{Bergeretal2003}). Using the detected core-collapse supernovae listed in \Tab~\ref{tab:SNRflux}, we derived a more accurate luminosity distribution at 5~GHz shown in \Fig~\ref{fig:snehist}. The data are divided into bins of size $\Delta$log$_{10}$(L$_{5;CasA}$\footnote{L$_{5;CasA}$ is the RSNe 5~GHz peak luminosity ratio to Cas~A's}) = 1. The dashed line presents the best fit Gaussian with mean luminosity ratio of 245.5, standard deviation of 1.27 and $\chi^{2}$ = 0.018. This distribution is similar to that presented by \citet{Lienetal2011}. 

However, this analysis ignores the upper limits (\Tab~\ref{tab:SNRupperlims}) entirely, and so we should consider whether these offer any significant additional information.
While their true luminosities are unknown, we need to determine if, given the upper limits, these supernovae are distributed differently to \Fig~\ref{fig:snehist}. To examine this we calculate the probability (or p-value), using the cumulative distribution function derived from the Gaussian distribution in \Fig~\ref{fig:snehist}, that a supernova will have a luminosity less than the upper limits listed in \Tab~\ref{tab:SNRupperlims}. The  p-values (\Tab~\ref{tab:SNRupperlims}), although on average are lower than 0.5, are probably consistent with the chosen luminosity function given that the upper limits may not have been observed at a time corresponding to when the peak flux density actually occurs. However, detailed work on testing whether the upper limits in \Tab~\ref{tab:SNRupperlims} are consistent with the chosen model are beyond the scope of this paper. 

\subsubsection*{Improved spectral indices}

\citet{Weileretal1986,Weileretal2007} showed that the spectral index between 6~cm and 20~cm evolves with time, for a number of RSNe. This has also been shown to be true for SN~1993J for wavelength pairs of 1.2/2~cm, 2/3.6~cm and 3.6/6~cm \citep{Weileretal2007}. Thus, it is expected that 2.3/5~GHz, 5/8.3~GHz, 5/15~GHz and  5/23~GHz spectral indices will display similar variations with time. To account for this, we have included light curves that best describe the spectral indices between 6 $\&$ 20~cm from SN~1980k \citep{Weileretal1986} and between 6 $\&$ 3.6~cm, 6 $\&$ 2~cm, 6 $\&$ 1.2~cm from SN~1993J \citep{Weileretal2007}. 

\subsubsection*{The new supernova rate upper limit}

\Tab~\ref{tab:SNrate} lists the results of applying the improved model parameters described above and listed in column 4 of \Tab~\ref{tab:SNRmodelparams}, to the Monte-Carlo simulation of LT06 (see \Sect~3.3.2). We find that the proportion of SNRs detected, $\beta_{SN}$ at the end of each epoch are relatively consistent, with minor variations between 0.66 and 0.84. The major changes occur during epochs where the sensitivity varies by almost an order of magnitude. The consistently high $\beta_{SN}$ resulted in an upper limit on the supernova rate of $\nu_{SN}$ = 0.2~yr$^{-1}$ at the end of the final epoch. The results of the simulation indicate that the radio observations are detecting 60$\%$ - 80$\%$ of all SNRs in NGC~253, thus pointing to the existence of a small population of undetected SNRs. Moreover, the fact that no RSNe or SNRs has been observed in NGC~253 during 21 years of radio observation suggests a low rate of RSNe production in NGC~253.

%---------------------------------------------------------------------------------------
 
\begin{center}
\begin{table}[htp]\scriptsize
\begin{minipage}{8cm}
\renewcommand\footnoterule{}
\caption{\textsc{Supernova Rate Upper Limits from Monte Carlo Simulations}}
\label{tab:SNrate}
\medskip
\hfill{}
\begin{tabular}{cccccc}
\hline
\hline
 
 Epoch  & Time$^{a}$ & $\nu^{b}$ & Sensitivity$^{c}$ &$\beta_{SN}$ $^{d}$ & $\nu_{SN}$ $^{e}$ \\ 
 
& (yr) & (GHz) & (mJy) & - & (yr$^{-1}$)  \\ 
\hline
1 & ... & 5.0 & ... & ... &   \\ 
2 & 1.5 & 5.0 & 3.0 &  0.746 & $<$2.71 \\ 
3 & 2.5 & 5.0 & 3.0 &  0.713 & $<$1.04   \\ 
4 & 4.0 & 5.0 & 3.0 &  0.683 & $<$0.53   \\ 
5 & 4.0 & 8.3 & 0.3 &  0.803 & $<$0.34   \\ 
6 & 0.25& 15 & 0.6 &  0.838 &  $<$0.33   \\
7 & 4.5 & 2.3 & 1.2 &  0.697 & $<$0.25  \\ 
8 & 0.5 & 23  & 2.0 &  0.661 &  $<$0.24  \\ 
9 & 1.7 & 2.3 & 2.0 &  0.676 & $<$0.22 \\ 
10 & 0.8 & 2.3 & 2.2 &  0.689 & $<$0.21  \\ 
11& 1.0 & 2.3 & 0.9 & 0.791 & $<$0.20  \\
 
 \hline\\
 \multicolumn{6}{l}{$^{a}$Time between epochs}\\
  \multicolumn{6}{l}{$^{b}$Observed frequency}\\
  \multicolumn{6}{l}{$^{c}$The 5$\sigma$ sensitivity of the observations}\\
 \multicolumn{6}{l}{$^{d}$Proportion of RSNe detected at the end of each epoch.}\\
 \multicolumn{6}{l}{$^{e}$Supernovae rate based upon all observations prior to and } \\
  \multicolumn{6}{l}{including that epoch.}\\
\end{tabular}
\hfill{}
\vspace{-2ex}
\end{minipage}
\end{table}
\end{center}

The final result for $\nu_{SN}$ agrees with estimates determined from near infrared (NIR) observations of the [FeII] line (0.24~yr$^{-1}$) by \citet{Rosenbergetal2013}. Observations of the [FeII] line have been shown to be a strong tracer of shocks associated with SNRs \citep{Rosenbergetal2012, Rosenbergetal2013}. Dust grains in the interstellar medium containing Fe atoms are destroyed via the shocks. The process releases the Fe atoms, which are then ionized by the interstellar radiation field. In the post shock region, Fe$^{+}$ is excited by electron collisions, causing it to emit at NIR wavelengths.  However, there are possible contributions to the [FeII] line from shocks due to other processes such as mergers \citep{Rosenbergetal2012}. Thus, it is possible that the NIR [FeII] observations are overestimating the supernova rate upper limit. 

%Recent observations by \citet{Bolattoetal13} with ALMA\footnote{the Atacama Large Millimeter Array} may agree with the latter scenario. It has been shown previously that NGC~253 possesses a molecular wind, possibly driven by the local starburst \citep{}. In their observations, \citet{Bolattoetal13} discovered that the extraplanar molecular gas resulting from  the molecular winds, traces H$\alpha$ filaments, connected to expanding molecular shells within the starburst region. From their observations they measured the molecular outflow rate to be between 1 to 3 times the star-formation, indicating that starburst-driven winds may be limiting the star-formation activity and consequently the production of supernovae.

The major uncertainty in our model arises from the RSNe peak luminosity distribution.  First, the detection of RSNe has primarily been through follow-up observations of optical supernovae, resulting in a small and incomplete sample, whose luminosity depends on the sensitivity of the survey. Second, there is the possibility that many RSNe are not visible in the optical due to extinction and we may be looking at a slightly different sub-population
%and third, this distribution do not take into account the effects of environment although it plays a major role in the peak brightness of RSNe \citep{Weileretal2002}.

Further uncertainty comes from the assumed rise time of the RSNe at the different frequencies. In this model, for simplicity, constant rise times (the time to reach maximum radio luminosities) were assumed for the hypothetical RSNe at the different frequencies. However, this has been observed to be variable for different RSNe (e.g. \citealt{Weileretal1986,Weileretal2002} ). Additionally, the relation between the rise time and peak luminosity has also been observed to be variable \citep{CFN06} and does not necessarily follow a one-to-one relationship as assumed in the model presented.

Further refinement to the supernova rate of NGC~253 could be obtained from either a large unbiased sample of the peak flux densities of RSNe, or direct observations of the RSNe population in NGC~253. Both possibilities can be provided by the \textit{Square Kilometre Array}\footnote{http://www.skatelescope.org/}, (SKA) and new generation telescopes, such as the Karl G. Jansky VLA\footnote{https://science.nrao.edu/facilities/vla} (JVLA). The sensitivity of both instruments may provide the opportunity to detect a population of weak SNRs that are faint due to either free-free absorption or intrinsic properties. The SKA is expected to have a Type II supernova detection rate of $\sim$620 yr$^{-1}$~deg$^{-2}$ out to a redshift $z~\sim$~5 \citep{Lienetal2011}. The JVLA and precursors to the SKA may provide a detection rate of $\sim$160 yr$^{-1}$~deg$^{-2}$ out to  redshift $z~\sim$~3 \citep{Lienetal2011}.

 %While the exact nature of the Gaussian distribution (i.e. $\mu$ and $\sigma$) is dependent on the choice of binning, the overall change to $\nu_{SN}$ with different binning was found to be negligible. However, using a uniform distribution between the faintest and brightest detected RSNe (i.e. 2 - 13000 times Cas~A) for the peak luminosities, results in $\beta_{SN}~\sim$~99$\%$ and subsequently reduces $\nu_{SN}$ to 0.15~yr$^{-1}$

 %----------------------------------------------------------------------------------------------------------------------------------------------------------------------------------
\subsection{Star Formation Rate}
\label{sec:SFR}

With new upper limits on the supernova rate, we can provide a new estimate for the star formation rate (SFR) in NGC~253. Following \citet{Condon92}, the SFR can be determined directly from the supernova rate (and vice versa). By assuming that all stars with mass $>8M_{\odot}$ eventually form supernovae, the relation between SFR and the supernova rate is given by:

\begin{equation}
\left[\dfrac{SFR(M\geq5~M_{\odot})}{M_{\odot}~\mathrm{yr}^{-1}}\right] = 24.4\left[\dfrac{\nu_{SN}}{\mathrm{yr}^{-1}}\right]
\end{equation}

Using the new supernova rate upper limits of $\nu_{SN}<0.2$~yr$^{-1}$ (\Sect~\ref{sec:SR}), we find 
$SFR(M\geq5~M_{\odot})<4.9~M_{\odot}$~yr$^{-1}$. We can compare this upper limit to estimates from independent methods. 
If one assumes that the main contribution to the observed radio luminosity in a starburst are from RSNe, SNRs and thermal H~\textsc{ii} regions, then the SFR of that region is directly proportional to its radio luminosity at the observed radio wavelength and can be calculated using the relation given by \citet{Condon92} and \citet{Haarsmaetal2000}. \citet{Ottetal2005} observed a total flux density of 0.59 Jy for NGC~253 at 24~GHz. Using the relation by  \citet{Haarsmaetal2000} and the data from \citet{Ottetal2005} we obtain $SFR(M\geq5~M_{\odot})$ = 9.3$~M_{\odot}$~yr$^{-1}$. Additional observations of NGC~253 at 23~GHz (total flux density of 0.56~Jy) by \citet{Takanoeal2005} give a similar $SFR(M\geq5~M_{\odot})$ = 8.7$~M_{\odot}$~yr$^{-1}$ for the $\sim$300~pc nuclear region, using the same method.

The $SFR(M\geq5~M_{\odot})$ can also be determined from far infrared (FIR) emission using equation 26 from \citet{Condon92}. From FIR luminosity measurements with \textit{IRAS}\footnote{http://irsa.ipac.caltech.edu/IRASdocs/exp.sup/ch1/index.html} \citep{Radovichetal2001} and scaling for the new distance 3.44~Mpc, the SFR is 1.3 - 2.0 $~M_{\odot}$~yr$^{-1}$ for the inner $\sim$300~pc nuclear region and 2.6 - 3.2 $~M_{\odot}$~yr$^{-1}$ for the entire galaxy. FIR luminosity measurements of the inner $\sim$350~pc nuclear region by \textit{Spitzer}\footnote{http://www.spitzer.caltech.edu/} \citep{PA2012} lead to an estimate of 2.7$~M_{\odot}$~yr$^{-1}$. It is encouraging that the $SFR(M\geq5~M_{\odot})$ calculated using different estimators are in broad agreement with each other. However, the differences point to the existence of systematic or intrinsic differences between the different estimators.

The supernova rate derived from the FIR luminosity (via \eqn~20 from  \citealt{Condon92}) is 0.05 - 0.13~yr$^{-1}$and  is similar to the lower limit of the supernova rate deduced from expansion rates (LT06). 
%and $\sim~40\%$ lower than the upper limit obtained from the Monte-Carlo simulations using the Gaussian distributed peak luminosities as described in \Sect~\ref{sec:SR}. 
 %----------------------------------------------------------------------------------------------------------------------------------------------------------------------------------
\section{Summary}
\label{sec:sum}

We have presented the results of multi-epoch observations of the southern starburst galaxy, NGC~253 with the LBA at 2.3~GHz. The results presented here are complementary to previous 2.3~GHz LBA observations by \citet{LT06}. We find the following results:

\begin{enumerate}

\item Seven compact sources were detected in the highest sensitivity observation (the 2008 epoch). All sources were identified with higher-frequency VLA observations (UA97), while six were identified with a 2.3~GHz LBA detections by LT06. The three brightest sources were also detected in the lower sensitivity observations (2006 and 2007).

\item The shell-like SNR, 5.48-43.3, was successfully imaged with the highest resolution in all three epochs. The structure in the 2008 image shows a double-lobed morphology which is dominated by the eastern lobe. The weaker, western lobe was not detected in the 2006 and 2007 observations, which is attributed to lower sensitivity and fidelity, due the lack of the 70~m Tidbinbilla antenna.

\item Observed differences in the small-scale structure of  5.48$-$43.3 between the 2004 (LT06) and the 2008 (this paper) observations can be explained by deconvolution errors associated with sparse $uv$-plane sampling.  No discernible expansion has been observed for 5.48$-$43.3 between the 2004 and 2008 epochs. 

\item The spectra of the 20 compact sources in NGC~253 from LT06 are compared with the spectra obtained with the new 2.3~GHz LBA flux density measurements (2008 epoch). As with LT06, we find that the spectra fit a free-free absorption model, with little difference in the spectra between the epochs.

\item Our results show no change in the classification of the compact sources by LT06. 12 of the 20 sources have steep spectra associated with SNRs, while the remaining eight have flat intrinsic power-law spectra ($\alpha>0.4$), indicative of H~\textsc{ii} regions. 

%\item Based upon the lack of variability and expected rise time for RSNe given the flux density of the compact objects in NGC~253, we conclude that the compact sources are consistent with very old SNRs affected by foreground free-free absorption.

\item We derive an improved RSNe peak luminosity distribution at 5~GHz using data from the  literature for 51 core-collapse supernovae (29 Type~II $\&$ 22 Type~Ibc).

\item We estimate a value for the upper limit of the supernova rate of  0.2~yr$^{-1}$ for the inner 300~pc of NGC~253, using an improved model that is based upon the principles of UA91 and LT06. This result was found to be consistent with estimates from near infrared (NIR) observations.

\item The results of our model indicate that the radio observations are detecting 60$\%$ - 80$\%$ of all SNRs in NGC~253, thus pointing to the existence of a small population of undetected SNRs. Moreover, no RSNe or SNRs has been observed in NGC~253 during 21 years of radio observation suggests a low rate of RSNe production in NGC~253.

%The model took into consideration: the mean free-free opacity of LT06; the sensitivity limits of 11 observations over a period of a 20.6 years;  the improved RSNe luminosity distribution and an improved distance estimate. To include multi-wavelength radio observations, we make use of the light-curves from \citet{Weileretal1986, Weileretal2007} to incorporate time-dependent spectral indices into our model.

\item A new upper limit to the star formation rate of $SFR(M\geq5~M_{\odot})<4.9~M_{\odot}$~yr$^{-1}$ is estimated directly from the supernova rate limits for the inner 300~pc region of the galaxy, which is consistent with independent estimates.

\end{enumerate}

 %----------------------------------------------------------------------------------------------------------------------------------------------------------------------------------
\acknowledgments

The International Centre for Radio Astronomy Research is a joint venture between Curtin University and the University of Western Australia, funded by the state government of Western Australia and the joint venture partners. The Long Baseline Array is part of the Australia Telescope which is funded by the Commonwealth of Australia for operation as a National Facility managed by CSIRO. SJT is a Western Australian Premier's Research Fellow, funded by the state government of Western Australia. The Centre for All-sky Astrophysics (CAASTRO) is an Australian Research Council Centre of Excellence, funded by grant CE110001020. We wish to thank Paola Castangia $\&$ Andreas Brunthaler for use of their VLA data, Cormac Reynolds for assisting with the calibration of the LBA data and Cath Trott for the helpful discussions on statistics. This research has made use of material from the Bordeaux VLBI Image Database (BVID) and the the NASA/IPAC Extragalactic Database (NED) which is operated by the Jet Propulsion Laboratory, California Institute of Technology, under contract with the National Aeronautics and Space Administration. We thank the anonymous referee for comments that improved this paper.

{\it Facilities:} \facility{ATCA, Mopra, Parkes, Ceduna, Hobart, Tidbinbilla}

%\clearpage

 %----------------------------------------------------------------------------------------------------------------------------------------------------------------------------------

%---------------------------------------------------------------------------------------------

\end{document}